\documentclass[11pt]{article}

\usepackage{epsfig}
\begin{document}

\baselineskip=14pt plus 0.2pt minus 0.2pt
\lineskip=14pt plus 0.2pt minus 0.2pt

\newcommand{\hlf}{\mbox{$1\over2$}}
\newcommand{\lfrac}[2]{\mbox{${#1}\over{#2}$}}
\newcommand{\scsz}[1]{\mbox{\scriptsize ${#1}$}}


\begin{flushleft}
\Large{\bf The Schr\"odinger system 
${H}=-\lfrac{1}{2}e^{\Upsilon(t-t_o)}\partial_{xx} 
+\lfrac{1}{2}\omega^2e^{-\Upsilon(t-t_o)}x^2$}
\vspace{0.25in}

\large
\bigskip

Michael Martin Nieto${\dagger}$ and D. Rodney Truax${\ddagger}$ \\
\normalsize
\vskip 10pt
${\dagger}$Theoretical Division (MS-B285), Los Alamos National Laboratory,
University of California, 
Los Alamos, New Mexico 87545, U.S.A. \\
${\ddagger}$Department of Chemistry,
 University of Calgary,
Calgary, Alberta T2N 1N4, Canada \\
\vskip 10pt
${\dagger}$Email:  mmn@lanl.gov \\
${\ddagger}${Email:  truax@ucalgary.ca} \\

\vskip 20pt
\today

\vspace{0.3in}
 
\end{flushleft}

\baselineskip=.33in

\noindent {\bf Abstract.}  
In this paper, we attack the specific time-dependent 
Hamiltonian problem 
${H}=-\lfrac{1}{2}e^{\Upsilon(t-t_o)}\partial_{xx} 
+\lfrac{1}{2}\omega^2e^{-\Upsilon(t-t_o)}x^2$.
This corresponds to a time-dependent mass ($TM$) Schr\"odinger
equation.  We give the specific transformations to i) the more 
general quadratic ($TQ$) Schr\"odinger equation and to ii) a different 
time-dependent oscillator ($TO$) equation.  
For each Schr\"odinger system, 
we give the Lie algebra of space-time symmetries, 
the number states, the coherent states, 
the squeezed-states and the time-dependent $\langle x\rangle$,  
$\langle p\rangle$, 
$(\Delta x)^2$, $(\Delta p)^2$, and  uncertainty product.

\vspace{1.25in}


\newpage

\normalsize

\baselineskip=.33in

\section{Introduction}

In recent work \cite{paperI,paperII}, we discussed 
general time-dependent quadratic ($TQ$) Schr\"odinger equations 
of the form 
\begin{eqnarray}
S_1\Phi(x,t) & = & \{-\left[1+k(t)\right]P^2+2T+h(t)D+
                       g(t)P    \nonumber\\
               &   & \hspace{1.0cm}-2h^{(2)}(t)X^2-2h^{(1)}(t)X
                    -2h^{(0)}(t)I\}\Phi(x,t) = 0. \label{e:pre12}
\end{eqnarray}
It was shown how to solve them by  i) first performing a
unitary transformation to a time-dependent mass ($TM$) equation, 
and then ii) making a change of time variable to yield a 
time-dependent oscillator
($TO$) equation which in principle can be solved.  One can then work  
backwards to find the $TM$ and $TQ$ solutions.

Elsewhere \cite{TMI}, 
we went into detail on how to solve a specific subclass 
of cases, $TM$ equations with only time-dependent $P^2$ and $X^2$ 
terms.  [We will refer to equations from this paper as, e.g., Eq. 
(\cite{TMI}-22).]
These Hamiltonians are parametrized as  
\begin{equation}
\hat{H}_2=\lfrac{1}{2}e^{-2\nu(t)}P^2+h^{(2)}(t)e^{2\nu(t)}X^2.
\label{genH2}
\end{equation} 
In this paper, we will demonstrate this procedure by examining 
the  $TM$ system 
\begin{equation}
\hat{H}_2=\lfrac{1}{2}e^{\Upsilon(t-t_o)}P^2 
          +\lfrac{\omega^2}{2}e^{-\Upsilon(t-t_o)}X^2. \label{oureq}
\end{equation}    

Starting with foresight with what we know to be the associated 
$TQ$ equation \cite{paperI}, this problem is:
\begin{eqnarray}
TQ:~~~~~~~~~~~&
S_1\Phi(x,t)=\left\{-P^2+2T+\Upsilon D-\omega^2X^2\right\}
\Phi(x,t)=0,& \label{e:hu1a} \\
&H_1 = \lfrac{1}{2}P^2-\lfrac{\Upsilon}{2} D +\lfrac{\omega^2}{2}X^2.& 
\end{eqnarray}
Note that this $TQ$ equation is actually time-independent.

This $TQ$ equation yields a $TM$ equation by the unitary transformation
\begin{equation}
R(0,\nu,0)=\exp\left\{-i\nu D\right\} = 
\exp\left\{-i\lfrac{1}{2}\Upsilon(t-t_o) D\right\},\label{tr1}
\end{equation}
where $\nu$ is a real function of $t$. This yields:  
\begin{eqnarray}
TM:~~~~~~&\hat{S}_2\hat{\Theta}(x,t)=\left\{-e^{\Upsilon(t-t_o)}P^2
+2T-\omega^2e^{-\Upsilon(t-t_o)}X^2\right\}\hat{\Theta}(x,t)
=0,&  \label{e:hu1b}
\end{eqnarray}
from which the $TM$ Hamiltonian of Eq. (\ref{oureq}) follows. 
This $TM$ equation is the defining equation of our problem. 
On its own, it has been the object of a number of investigations
\cite{paperI}, \cite{ek1}-\cite{ps1}.

The $TM$ equation yields a $TO$ equation by the change of 
time variable \cite{paperI}
\begin{equation} 
t'-t_o'=\frac{1}{\Upsilon}\left[e^{\Upsilon(t-t_o)}-1\right].
\label{e:hu8}
\end{equation}
One finds: 
\begin{eqnarray}
TO:~~~~~~~~~&
S_3\Psi(x,t')=\left\{-P^2+2T'-{{\omega^2}\over{\left[1+\Upsilon
(t-t_o)\right]^2}}X^2\right\}\Psi(x,t')=0, &  \label{e:hu1c}  \\
& H_3 = \frac{1}{2}P^2+{{\omega^2}\over{2[1+\Upsilon(t-t_o)]^2}}X^2. &
\end{eqnarray}
Depending upon the sign of $\Upsilon$, certain restrictions upon 
$t'$ apply.  If $\Upsilon > 0$, then $t'-t_o'$ will lie in the interval 
$[0,+\infty)$.  If $\Upsilon < 0$, $t'-t_o'$ must lie in the 
interval $[0,1/|\Upsilon|)$. 
Because i) three equations ($TQ,~TM,~TO$) 
are being considered, ii) there is a condition on the relative 
sizes of $|\Upsilon|^2$ and $\omega^2$, and iii) the sign of 
$\Upsilon$is important, there are 18 separate cases that can be discussed 
(6 for each Schr\"odinger equation). 
In what now follows we will look at specific cases for 
each equation as illustrative examples.  

In Section 2, we derive time-dependent functions from which   one can 
compute the time-dependent Lie symmetries that form the basis of 
the oscillator algebras.  
In Sections  4, 5, and 6, number states, 
coherent states, and squeezed states are obtained.   
In Section 7, we obtain coherent-state and squeezed-state 
expectation values, uncertainties, and uncertainty 
products for the three systems, and discuss the classical motion.


\section{Solutions for the Time-Dependent Functions}

\subsection{The $TO$ Functions}

To find the symmetries associated with the $TO$ equation,
$S_3\Psi=0$,  we must obtain the complex solutions $\xi$ 
and $\bar{\xi}$ to the differential Eq. (\cite{TMI}-19).  For our problem, 
this equation has the form
\begin{equation}
\ddot{\gamma}+{{\omega^2}\over{\left[1+\Upsilon
(t'-t_o')\right]^2}}\gamma=0.\label{e:hu12}
\end{equation}
The solutions $\xi$ and their complex conjugates $\bar{\xi}$ satisfy the 
Wronskian (\cite{TMI}-20)
\begin{equation}
W_{t'}(\xi,\bar{\xi})=-i, \label{e:hu12cw}
\end{equation}
where subscript  indicates that 
$\xi$ and $\bar{\xi}$ are to be differentiated with respect to  $t'$.  

Make the following change of variables in Eq. (\ref{e:hu12}): 
\begin{equation}
\tau=1+\Upsilon(t'-t'_o),~~~~w(\tau)=\gamma(t).\label{e:hu14}
\end{equation}
With this change of variables, Eq. (\ref{e:hu12}) becomes
\begin{equation}
\frac{d^2w}{d\tau^2}+\frac{\omega^2/\Upsilon^2}{\tau^2}w=0.
\label{e:hu12a}
\end{equation}
When $\Upsilon > 0$, then $\tau\in [1,\infty)$.  
When $\Upsilon < 0$, then $\tau\in (0,1]$.  In either case, $\tau$ 
is always positive.

The general solutions to Eq. (\ref{e:hu12a}) are derived in Appendix A.  
There are three different classes of solutions, 
depending upon the relative magnitudes of $\Upsilon^2$ and $\omega^2$.  
The three classes of real solutions to Eq. (\ref{e:hu12a}) are:
\begin{eqnarray}
1.~~\Upsilon^2 > 4\omega^2~~~~~~~~~~
w_1(\tau)=\sqrt{\tau}\exp\left(-\lfrac{\Delta}{2}\ln{\tau}\right),&~~~~&
w_2(\tau)=\sqrt{\tau}\exp\left(\lfrac{\Delta}{2}\ln{\tau}\right).
\label{e:hu12ars>} \\
 2.~~ \Upsilon^2 = 4\omega^2~~~~~~~~~~~~~~~~~~~~~~~~~~~~~~
w_1(\tau)=\sqrt{\tau},&~~~~&w_2(\tau)\sqrt{\tau}\ln{\tau}.
\label{e:hu12ars=} \\
3.~~\Upsilon^2 < 4\omega^2~~~~~~~~~~~~~
w_1(\tau)=\sqrt{\tau}\cos\left({\lfrac{\Delta}{2}\ln{\tau}}\right),&~~~~&
w_2(\tau)=\sqrt{\tau}\sin\left({\lfrac{\Delta}{2}\ln{\tau}}\right),
\label{e:hu12ars<} \\
\Delta^2  \equiv  \left|1-4\omega^2/\Upsilon^2\right|. & &
\label{e:hu16x}
\end{eqnarray}

Real solutions to Eq. (\ref{e:hu12}) are obtained from these solutions by 
combining Eqs. (\ref{e:hu14}) and (\ref{e:hu12ars>}) to (\ref{e:hu12ars<}).  
We write the solutions as
\begin{equation}
\gamma_1(t')=C_1w_1(\tau),~~~~~~~
\gamma_2(t')=C_2w_2(\tau).\label{e:hu12r}
\end{equation}   
The constants $C_1$ and $C_2$ are chosen so that the Wronskian of the 
real solutions is
\begin{equation}
W_{t'}(\gamma_1,\gamma_2)=\gamma_1\dot{\gamma}_2-\dot{\gamma}_1\gamma_2
=1.\label{e:hu12rw}
\end{equation}
Both $C_1$ and $C_2$ will depend upon the magnitude and sign of $\Upsilon$ 
and this gives rise to two subclasses of solutions.  
For example, for class-1 solutions, when $\Upsilon>0$, we find that 
\begin{equation}
C_1=\sqrt{\frac{1}{\Upsilon\Delta}}=C_2.  \label{e:hu12C12>}
\end{equation}
\noindent For class-2 solutions, 
when $\Upsilon<0$ ($\Upsilon=-|\Upsilon|$), then
\begin{equation}
C_1=\sqrt{\frac{1}{|\Upsilon|\Delta}}=-C_2.  \label{e:hu12C12<}
\end{equation}

The complex solutions satisfying the Wronskian 
(\ref{e:hu12cw}) are 
\begin{equation}
\xi(t')=\sqrt{\lfrac{1}{2}}\left(\gamma_1(t')+i\gamma_2(t')\right),
\label{e:hu12cs}
\end{equation}
and their complex conjugates, $\bar{\xi}(t')$.
Citing the class-1 and class-2 examples again, 
we obtain  solutions depending upon the sign of $\Upsilon$.  
For class-1 ($\Upsilon>0$)  we have
\begin{equation}
\xi(t') = \sqrt{\lfrac{1}{2\Upsilon\Delta}}\left(e^{-\lfrac{\Delta}{2}
\ln{\tau}}+ie^{\lfrac{\Delta}{2}\ln{\tau}}\right),\label{e:hu12cY>0}
\end{equation}
and for class-2 ($\Upsilon <0$)  
\begin{equation}
\xi(t')=\sqrt{\lfrac{1}{2|\Upsilon|\Delta}}\left(e^{-\lfrac{\Delta}{2}
\ln{\tau}}-ie^{\lfrac{\Delta}{2}\ln{\tau}}\right).\label{e:hu12cY<0}
\end{equation}

In order to compute the $os(1)$ generators, expectation values, uncertainties, 
and uncertainty products associated with the $TO$ Schr\"odinger equation, 
several other functions are required.  These 
functions, $\phi_j$, $j=1,2,3$, [see Eq. (\cite{TMI}-20)], and their first 
and second derivatives can all be calculated from $\xi$ and 
$\bar{\xi}$ and their derivatives.  They are given in Appendix B, Table B-1.


\subsection{The $TM$ Functions}

The $TM$ functions, 
$\hat{\xi}$, $\hat{\bar{\xi}}$, $\hat{\dot{\xi}}$, $\hat{\dot{\bar{\xi}}}$, 
$\hat{\phi_j}$, $\hat{\dot{\phi}}_j$, $\hat{\ddot{\phi}}_j$, 
$j=1,2,3$, can be obtained from the composition of the corresponding 
$TO$ functions and the mapping of Eq. (\ref{e:hu8}).  
[See Eq. (\cite{TMI}-32).]  
These $TM$ functions are compiled in Appendix B, Table B-2.


\subsection{The $TQ$ Functions}

The $TQ$-functions, $\Xi_P$ and $\Xi_X$, their complex conjugates, 
and $C_{3,T}$, $C_{3,D}$, and $C_{3,X^2}$   can 
be computed from the $TM$ functions in Table B-2 using Eqs. (\cite{TMI}-41) 
to (\cite{TMI}-44).  In particular, 
\begin{equation}
\Xi_P(t)=\hat{\xi}(t)e^{\nu}=\hat{\xi}(t)e^{-\chi/2},~~~~
\Xi_X(t)=\hat{\xi}(t)e^{-\nu}=\hat{\dot{\xi}}(t)e^{\chi/2}
\label{e:hu12tqc1}
\end{equation}
since $\kappa=0$ and  where $\chi=\Upsilon(t-t_o)$.
The $TQ$-functions are listed in Appendix B, Table B-3.

\subsection{Initial Values of Functions}

Initial values for some of these functions are also needed.  For $TO$, the 
initial values are obtained by setting $t'=t_o'$ in the functions in 
Table B-1.  Similarly, for $TM$ and $TQ$, the initial values result
by setting $t=t_o$ in the functions in Tables B-2 and B-3.  When 
$t'=t_o'$ for $TO$, 
the variable $\tau=1$, regardless of the sign of $\Upsilon$.  
When $t=t_o$ for $TM$ and $TQ$, the variable $\chi=0$.


\section{The Algebras}

Although each class of solutions will produce a Schr\"odinger algebra, 
(${\cal SA}^c_1$), we are only interested in its oscillator subalgebra 
$os(1)$.  So, the three Schr\"odinger equations 
will have six isomorphic $os(1)$  algebras: three corresponding 
to $\Upsilon > 0$ and three to $\Upsilon < 0$, for a total of 18 
isomorphic Lie algebras.  We use the following notation to distinguish 
between the eighteen different cases, i.e., systems: 

{\bf Notation:} $\{TX, R(\Upsilon,\omega), \pm\}$, 
where $TX=\{TO,TM,TQ\}$, $R(\Upsilon,\omega)$ specifies one of 
the three 
relationships between $\Upsilon^2$ and $4\omega^2$, and $\pm$ indicates 
the sign of $\Upsilon$. 

The basis operators for the $os(1)$ algebras, 
$\{M_j,J_{j\pm}\}$, can be constructed from the appropriate functions in 
Tables B-1, B-2, and B-3, using  
the third operator in Eq. (\cite{TMI}-17) and Eq. (\cite{TMI}-18) for $TO$,  
Eqs.(\cite{TMI}-30) and (\cite{TMI}-31) for $TM$, and  Eq. (\cite{TMI}-40) 
for the $TQ$ system.  
There are too many algebras to list the operators of each.  So, as 
examples, we give the operators for a few of the  systems 
$\{TX,R(\Upsilon,\omega),\pm\}$:

{\bf A Case of a $TO$  $os(1)$ Algebra:}
\begin{eqnarray}
\{TO,~\Upsilon^2<4\omega^2\,+\}:~~~~~~~~
J_{3-} & = & i\sqrt{\lfrac{1}{\Upsilon\Delta}}\exp{\left[i\lfrac{\Delta}{2}
\ln{\tau}\right]}\left\{\sqrt{\tau}P-\lfrac{\Upsilon}{2}(1+i\Delta)
\frac{1}{\sqrt{\tau}}X\right\},\nonumber\\*[1mm]
J_{3+} & = & i\sqrt{\lfrac{1}{\Upsilon\Delta}}\exp{\left[-i\lfrac{\Delta}{2}
\ln{\tau}\right]}\left\{-\sqrt{\tau}P+\lfrac{\Upsilon}{2}(1-i\Delta)
\frac{1}{\sqrt{\tau}}X\right\},\nonumber\\*[1mm]
M_3 & = & \lfrac{2}{\Upsilon\Delta}\tau T'-\lfrac{1}{\Delta}D,
\label{e:ex1op1}
\end{eqnarray}
where $\tau$ is given by Eq. (\ref{e:hu14}).

{\bf Cases of  $os(1)$ $TM$ Algebras:}
Recalling that $\chi=\Upsilon(t-t_o)$: 
\begin{eqnarray}
\{TM,\Upsilon^2<4\omega^2,+\}:~~~~~~~
\hat{J}_{2-} & = & i\sqrt{\lfrac{1}{\Upsilon\Delta}}e^{\chi/2}
e^{i\frac{\Delta}{2}\chi}\left\{P-\lfrac{\Upsilon}{2}(1+i\Delta)e^{-\chi}
X\right\},\nonumber\\*[1mm]
\hat{J}_{2+} & = & i\sqrt{\lfrac{1}{\Upsilon\Delta}}e^{\chi/2}
e^{-i\frac{\Delta}{2}\chi}\left\{-P+\lfrac{\Upsilon}{2}(1-i\Delta)e^{-\chi}
X\right\},\nonumber\\*[1mm]
\hat{M}_{2} & = & \lfrac{2}{\Upsilon\Delta}T-\lfrac{1}{\Delta}D.
\label{e:tmop<+} \\
\{TM,\Upsilon^2<4\omega^2,-\}:~~~~~~~
\hat{J}_{2-} & = & i\sqrt{\lfrac{1}{|\Upsilon|\Delta}}e^{\chi/2}
e^{-i\frac{\Delta}{2}\chi}\left\{P+\lfrac{|\Upsilon|}{2}(1-i\Delta)e^{-\chi}
X\right\},\nonumber\\*[1mm]
\hat{J}_{2+} & = & i\sqrt{\lfrac{1}{|\Upsilon|\Delta}}e^{\chi/2}
e^{i\frac{\Delta}{2}\chi}\left\{-P-\lfrac{|\Upsilon|}{2}(1+i\Delta)e^{-\chi}
X\right\},\nonumber\\*[1mm]
\hat{M}_{2} & = & \lfrac{2}{|\Upsilon|\Delta}T+\lfrac{1}{\Delta}D.
\label{e:tmop<-} \\
\{TM,\Upsilon^2=4\omega^2,+\}:~~~~~~~
\hat{J}_{2-} & = & i\sqrt{\lfrac{1}{2\Upsilon}}e^{\chi/2}
\left\{(1+i\chi)P-\Upsilon e^{-\chi}\left[\lfrac{1}{2}+i\left(
1+\lfrac{1}{2}\chi\right)\right]\right\},\nonumber\\*[1mm]
\hat{J}_{2+} & = & i\sqrt{\lfrac{1}{2\Upsilon}}e^{\chi/2}
\left\{-(1-i\chi)P+\Upsilon e^{-\chi}\left[\lfrac{1}{2}-i\left(
1+\lfrac{1}{2}\chi\right)\right]\right\},\nonumber\\*[1mm]
\hat{M}_{2} & = & \lfrac{1}{\Upsilon}(1+\chi^2)T-\lfrac{1}{2}(1+\chi)^2D
+\lfrac{1}{2}e^{-\chi}(1+\chi)X^2.\label{e:tmop=+} 
\end{eqnarray}
\noindent $\{TM,\Upsilon^2>4\omega^2,+\}$
\begin{eqnarray}
\hat{J}_{2-} & = & i\sqrt{\lfrac{1}{2\Upsilon\Delta}}e^{\chi/2}
\left\{\left(e^{-\frac{\Delta}{2}\chi}+ie^{\frac{\Delta}{2}\chi}\right)P
-\lfrac{\Upsilon}{2}e^{-\chi}\left[(1-\Delta)e^{-\frac{\Delta}{2}\chi}+i
(1+\Delta)e^{\frac{\Delta}{2}\chi}\right]X\right\},\nonumber\\*[1mm]
\hat{J}_{2+} & = & i\sqrt{\lfrac{1}{2\Upsilon\Delta}}e^{\chi/2}
\left\{-\left(e^{-\frac{\Delta}{2}\chi}-ie^{\frac{\Delta}{2}\chi}\right)P
+\lfrac{\Upsilon}{2}e^{-\chi}\left[(1-\Delta)e^{-\frac{\Delta}{2}\chi}-i
(1+\Delta)e^{\frac{\Delta}{2}\chi}\right]X\right\},\nonumber\\*[1mm]
\hat{M}_2 & = & \lfrac{1}{2\Upsilon\Delta}\left(e^{-\Delta\chi}
+e^{\Delta\chi}\right)T-\lfrac{1}{2\Delta}\left[(1-\Delta)e^{-\Delta\chi}
+(1+\Delta)e^{\Delta\chi}\right]D\nonumber\\*[1mm]
 &  & \hspace{2.5cm}-\lfrac{\Upsilon}{4}e^{-\chi}
\left[-(1-\Delta)e^{-\Delta\chi}+(1+\Delta)e^{\Delta\chi}\right]X^2.
\label{e:tmop>+}
\end{eqnarray}

{\bf A Case of an $os(1)$ $TQ$ Algebra:}
\begin{eqnarray}
\{TQ,~\Upsilon^2<4\omega^2,+\}:~~~~~~~
J_{1-} & = & i\sqrt{\lfrac{1}{\Upsilon\Delta}}\exp{\left(i
\lfrac{\Delta}{2}\chi\right)}\left[P-\lfrac{\Upsilon}{2}(1+i\Delta)X
\right],\nonumber\\*[1mm]
J_{1+} & = & i\sqrt{\lfrac{1}{\Upsilon\Delta}}\exp{\left(-i
\lfrac{\Delta}{2}\chi\right)}\left[-P+\lfrac{\Upsilon}{2}(1-i\Delta)X
\right],\nonumber\\*[1mm]
M_1 & = & \lfrac{2}{\Upsilon\Delta}T.\label{e:ex1op3}
\end{eqnarray}


\section{Number States}

Combining Section 4 of Ref. \cite{TMI} with  
the time-dependent functions for this problem  
given in Tables B-1 to B-3, one can 
exploit the $os(1)$ algebraic structure to obtain the  
number states for the 18 systems characterized by 
$\{TX,R(\Upsilon,\omega),\pm)$. We give examples below:  

{\bf A Case of $TO$ Number States:} 
For  the system 
$\{TO,\Upsilon^2<4\omega^2,+\}$, 
for which the Lie symmetry operators are given in Eq. (\ref{e:ex1op1}), 
the wave functions are 
\begin{eqnarray}
& \Psi_n(x,t)  =  \sqrt{\frac{1}{2^n n!}}
\exp{\left(\frac{i}{4}\Upsilon\frac{x^2}{\tau}\right)}
H_n\left(\sqrt{\frac{\Upsilon\Delta}{2}}\frac{x}{\sqrt{\tau}}\right)
  \left(\frac{\Upsilon\Delta}{2\pi\tau}
\right)^{\lfrac{1}{4}}\exp{\left[-\lfrac{i}{2}\left(n+\lfrac{1}{2}
\right)\Delta\ln{\tau}\right]}.&  \label{e:h62}
\end{eqnarray}
[See Eq. (\cite{TMI}-54).]  Here and below, the $H_n$ are Hermite 
polynomials.

{\bf A Case of $TM$ Number States:}
For  the system 
$\{TM,\Upsilon^2<4\omega^2,+\}$, 
for which the Lie symmetry operators are given in Eq. (\ref{e:tmop>+}), 
the wave functions are [see Eq. (\cite{TMI}-58)]
\begin{eqnarray}
& \hat{\Theta}_n(x,t)  =  \sqrt{\frac{1}{2^n n!}}
\exp{\left(\frac{i}{4}\Upsilon\frac{x^2}{e^{\chi}}\right)}
H_n\left(\sqrt{\frac{\Upsilon\Delta}{2}}\frac{x}{e^{\chi/2}}\right)
\left(\frac{\Upsilon\Delta}{2\pi e^{\chi}}\right)^{\lfrac{1}{4}}
\exp{\left[-\lfrac{i}{2}\left(n+\lfrac{1}{2}\right)
                \Delta {\chi}\right]}. & \label{tmnew}
\end{eqnarray}

{\bf A Case of $TQ$ Number States:} 
For the system 
$\{TQ,\Upsilon^2<4\omega^2,+\}$, 
for which the Lie symmetry operators are given in Eq. (\ref{e:ex1op3}), 
the wave functions are [see Eq. (\cite{TMI}-63)]
\begin{eqnarray}
& \Phi_n(x,t)  =  \sqrt{\frac{1}{2^n n!}}
\exp{\left(\lfrac{i}{4}\Upsilon x^2\right)}
H_n\left(\sqrt{\lfrac{\Upsilon\Delta}{2}}x\right)
\exp{\left(-\lfrac{\Upsilon\Delta}{4}x^2\right)}
\left(\lfrac{\Upsilon\Delta}{2\pi}
\right)^{\lfrac{1}{4}}\exp{\left[-\lfrac{i}{2}\left(n+\lfrac{1}{2}\right)
\Delta \chi \right]}. & \label{e:h90}
\end{eqnarray}


\section{Coherent States}

Combining the time-dependent functions for this problem  given in 
Tables B-1 to B-3 (this time) with Section 5 of Ref. \cite{TMI}, 
one can exploit the $os(1)$ algebraic structure to obtain the  
coherent  states for the 18 systems characterized by 
$\{TX,R(\Upsilon,\omega),\pm)$. We give examples below:  


{\bf A Case of $TO$ Coherent States:} 
For specific cases, we use 
the appropriate $t'$-dependent functions in Table B-1. 
For example, for the case $\{TO,\Upsilon^2<4\omega^2,+\}$, 
the coherent states are [see Eq. (\cite{TMI}-66)]
\begin{eqnarray}
{\Psi}_{\alpha}(x,t') & = & \left(\lfrac{\Upsilon\Delta}{2\pi}
\right)^{\frac{1}{4}}e^{-\frac{1}{4}(1+i\Delta)\ln{\tau}}
\exp\left\{-\lfrac{1}{2}
\left(\lfrac{\Upsilon\Delta}{2\tau}\right)  
\left[{x}    - {X}_3^+(\alpha)\right]^2   \right\}
           \nonumber\\
 &   & \times\exp\left\{i\left[
\left(\lfrac{\Upsilon}{4\tau}\right) x^2     +
\left(\lfrac{\Upsilon\Delta}{2\tau}\right) 
\left(  {x} - \lfrac{1}{2}{X}_3^+(\alpha) \right)
{X}_3^-(\alpha) \right]\right\},  \label{new1}
\end{eqnarray}
\begin{eqnarray}
{X}_3^+(\alpha) 
 & = & 
p_o
\left(\lfrac{2\sqrt{\tau}}{\Upsilon\Delta}\right)
\sin{\left(\lfrac{\Delta}{2}\ln{\tau}\right)}
+x_o  \left(\lfrac{\sqrt{\tau}}{\Delta}\right)
\left[
\Delta\cos{\left(\lfrac{\Delta}{2}\ln{\tau}\right)}
-\sin{\left(\lfrac{\Delta}{2}\ln{\tau}\right)}
\right],\label{new2} \\
{X}_3^-(\alpha) 
 & = & 
p_o
\left(\lfrac{2\sqrt{\tau}}{\Upsilon\Delta}\right)
\cos{\left(\lfrac{\Delta}{2}\ln{\tau}\right)}
-x_o  \left(\lfrac{\sqrt{\tau}}{\Delta}\right)
\left[
\cos{\left(\lfrac{\Delta}{2}\ln{\tau}\right)}
+\Delta\sin{\left(\lfrac{\Delta}{2}\ln{\tau}\right)}
\right].\label{new3}
\end{eqnarray}
Recall that $\tau=1+\Upsilon(t'-t_o')$.


{\bf A Case of $TM$ Coherent States:} 
For $\{TM,\Upsilon^2<4\omega^2,+\}$, we find from Eq. (\cite{TMI}-68) that
\begin{eqnarray}
\hat{\Theta}_{\alpha}(x,t) & = & \left(\lfrac{\Upsilon\Delta}{2\pi}
\right)^{\frac{1}{4}}e^{-\frac{1}{4}(1+i\Delta){\chi}}
\exp\left\{-\lfrac{1}{2}
\left(\lfrac{\Upsilon\Delta}{2e^{\chi}}\right)  
\left[{x}    - \hat{X}_2^+(\alpha)\right]^2   \right\}
           \nonumber\\
 &   & \times\exp\left\{i\left[
\left(\lfrac{\Upsilon}{4e^{\chi}}\right) x^2     +
\left(\lfrac{\Upsilon\Delta}{2e^{\chi}}\right) 
\left(  {x} - \lfrac{1}{2}\hat{X}_2^+(\alpha) \right)
\hat{X}_2^-(\alpha) \right]\right\},
\label{e:h112+}
\end{eqnarray}
\begin{eqnarray}
\hat{X}_2^+(\alpha) 
 & = & 
p_o
\left(\lfrac{2e^{\chi/2}}{\Upsilon\Delta}\right)
\sin{\left(\lfrac{\Delta}{2}\chi\right)}
+x_o  \left(\lfrac{e^{\chi/2}}{\Delta}\right)
\left[
\Delta\cos{\left(\lfrac{\Delta}{2}\chi\right)}
-\sin{\left(\lfrac{\Delta}{2}\chi\right)}\right],
\label{e:h112+a}  \\
\hat{X}_2^-(\alpha) 
 & = & 
p_o
\left(\lfrac{2e^{\chi/2}}{\Upsilon\Delta}\right)
\cos{\left(\lfrac{\Delta}{2}\chi\right)}
-x_o  \left(\lfrac{e^{\chi/2}}{\Delta}\right)
\left[
\cos{\left(\lfrac{\Delta}{2}\chi\right)}
+\Delta\sin{\left(\lfrac{\Delta}{2}\chi\right)}
\right].
\label{e:h112+b}
\end{eqnarray}
Recall that $\chi=\Upsilon(t-t_o)$.


{\bf A Case of $TQ$ Coherent States:} 
Coherent states for each $TQ$-system [Eq. (\cite{TMI}-74)] of 
this problem  can be computed from the functions in Table B-3.
For example, the DOCS wave function for $\{TQ,\Upsilon^2<4\omega^2,+\}$
is 
\begin{eqnarray}
\Phi_{\alpha}(x,t) & = & 
\left(\lfrac{\Upsilon\Delta}{2\pi}
\right)^{\frac{1}{4}} e^{-i\Delta{\chi}/4}
\exp\left\{-\lfrac{1}{2}
\left(\lfrac{\Upsilon\Delta}{2}\right)  
\left[{x}    - {X}_1^+(\alpha)\right]^2   \right\}
           \nonumber\\
 &   & \times\exp\left\{i\left[
\left(\lfrac{\Upsilon}{4}\right) x^2     +
\left(\lfrac{\Upsilon\Delta}{2e^{\chi}}\right) 
\left(  {x} - \lfrac{1}{2}{X}_1^+(\alpha) \right)
{X}_1^-(\alpha) \right]\right\},
\label{newX1}
\end{eqnarray}
\begin{eqnarray}
{X}_1^+(\alpha) 
 & = & 
p_o
\left(\lfrac{2}{\Upsilon\Delta}\right)
\sin{\left(\lfrac{\Delta}{2}\chi\right)}
+x_o  \left(\lfrac{1}{\Delta}\right)
\left[
\Delta\cos{\left(\lfrac{\Delta}{2}\chi\right)}
-\sin{\left(\lfrac{\Delta}{2}\chi\right)}\right],
\label{newX2} \\
{X}_1^-(\alpha)  
 & = & 
p_o
\left(\lfrac{2}{\Upsilon\Delta}\right)
\cos{\left(\lfrac{\Delta}{2}\chi\right)}
-x_o  \left(\lfrac{1}{\Delta}\right)
\left[
\cos{\left(\lfrac{\Delta}{2}\chi\right)}
+\Delta\sin{\left(\lfrac{\Delta}{2}\chi\right)}
\right].
\label{newX3}
\end{eqnarray}


\section{Squeezed States}

Finally, combining the time-dependent functions for this problem  
given in Tables B-1 to B-3 (this time) with Section 6  of Ref. \cite{TMI}, 
one can exploit the $os(1)$ algebraic structure to obtain the  
coherent  states for the 18 systems characterized by 
$\{TX,R(\Upsilon,\omega),\pm)$. We give examples below:  


{\bf A Case of $TO$ Squeezed States:} 
The squeezed-state wave functions for the 
$\{TO,\Upsilon^2<4\omega^2,+\}$ system are obtained from 
Eq. (\cite{TMI}-91) and the functions in Table B-1.  They have the form 
\begin{eqnarray}
\Psi_{\alpha,z}(x,t') & = & 
\left(\frac{1}{2\pi {Q}_3}\right)^{\frac{1}{4}}
\left(\frac
{e^{-i\frac{\Delta}{2}\ln{\tau}}
   +e^{i(\frac{\Delta}{2}\ln{\tau}-\theta)}\tanh{r}}
{e^{i\frac{\Delta}{2}\ln{\tau}}
   +e^{-i(\frac{\Delta}{2}\ln{\tau}+\theta)}\tanh{r}}
\right)^{\frac{1}{4}}
\exp\left\{-\frac{1}{4}\frac{1}{{Q}_3}
\left[x-{X}_3^+(\alpha,z)\right]^2\right\}
\nonumber\\
 &   & 
\times\exp
\left\{i\left[\frac{1}{4}\frac{{R}_3}{{Q}_3}x^2
+\frac{1}{2{Q}_3}
\left(x-\lfrac{1}{2}{X}_3^+(\alpha,z)\right){X}_3^-(\alpha,z)
\right]\right\},
\label{e:h176} \\
{Q}_3 & = & \frac{\tau}{\Upsilon\Delta}
\left(\cosh{2r}+\cos{(\Delta\ln{\tau}-\theta)}\sinh{2r}\right)
\label{e:h177}  \\
\frac{R_3}{Q_3} & = & \frac{\Upsilon}{\tau}
\frac{\cosh 2r 
+[cos(\Delta\ln\tau -\theta)-\Delta\sin(\Delta\ln\tau -\theta)]\sinh 2r}
{\cosh{2r}+\cos{(\Delta\ln{\tau}-\theta)}\sinh{2r}},
\end{eqnarray}
where $X_3^+(\alpha)$ and $X_3^-(\alpha)$ are given in Eqs. (\ref{new2})
and Eq. (\ref{new3}). Also, from Eq. (\cite{TMI}-94).  one  has 
\begin{eqnarray}
& X_3^{-}(\alpha,z)
=X_3^{-}(\alpha) \cosh{2r} +Y_3^-(\alpha,\theta)\sinh{2r}, &
\label{s94} \\
& Y_1^-(\alpha,\theta)  = p_o
\left(\lfrac{2\sqrt{\tau}}{\Upsilon\Delta}\right)
\cos\left(\lfrac{\Delta}{2}\chi-\theta\right)
-x_o\left(\lfrac{\sqrt{\tau}}{\Delta}\right) 
\left[\cos\left(\lfrac{\Delta}{2}\chi-\theta\right)
      -\Delta\sin\left(\lfrac{\Delta}{2}\chi-\theta\right)\right]. &
\label{toy}
\end{eqnarray}


{\bf A Case of $TM$ Squeezed States:} 
The squeezed-state wave functions for the 
 $\{TM,\Upsilon^2<4\omega^2,+\}$  system are obtained from 
Eq. (\cite{TMI}-96) and the functions in Table B-2.  They have the form 
\begin{eqnarray}
\hat{\Theta}_{\alpha,z}(x,t) & = & 
\left(\frac{1}{2\pi \hat{Q}_2}\right)^{\frac{1}{4}}
\left(\frac
{e^{-i\frac{\Delta}{2}\chi}
   +e^{i(\frac{\Delta}{2}\chi-\theta)}\tanh{r}}
{e^{i\frac{\Delta}{2}\chi}
   +e^{-i(\frac{\Delta}{2}\chi+\theta)}\tanh{r}}
\right)^{\frac{1}{4}}
\exp\left\{-\frac{1}{4}\frac{1}{\hat{Q}_2}
\left[x-\hat{X}_2^+(\alpha,z)\right]^2\right\}
\nonumber\\
 &   & 
\times\exp
\left\{i\left[\frac{1}{4}\frac{\hat{R}_2}{\hat{Q}_2}x^2
+\frac{1}{2\hat{Q}_2}
\left(x-\lfrac{1}{2}\hat{X}_2^+(\alpha,z)\right)\hat{X}_2^-(\alpha,z)
\right]\right\}, \\
\label{tmss}
\hat{Q}_2& = & \frac{e^{\chi}}{\Upsilon\Delta}
\left(\cosh{2r}+\cos{(\Delta\ln{\tau}-\theta)}\sinh{2r}\right)
\label{ssq2} \\
\frac{\hat{R}_2}{\hat{Q}_2} & = & \frac{\Upsilon}{e^{chi}}
\frac{\cosh 2r 
+[cos(\Delta\chi -\theta)-\Delta\sin(\Delta\chi -\theta)]\sinh 2r}
{\cosh{2r}+\cos{(\Delta\chi-\theta)}\sinh{2r}},
\end{eqnarray}
where $\hat{X}_2^+(\alpha)$ and where $\hat{X}_2^-(\alpha)$
are  given in Eqs. (\ref{e:h112+a}) and  (\ref{e:h112+b}). 
Also, from Eq. (\cite{TMI}-101),  
\begin{eqnarray}
\hat{X}_2^{-}(\alpha,z)
& = & \hat{X}_2^{-}(\alpha) \cosh{2r} +\hat{Y}_2^-(\alpha,\theta)\sinh{2r},
\label{ssx2-}  \\
\hat{Y}_2^-(\alpha,\theta) & = & 
p_o\left(\lfrac{2e^{\chi/2}}{\Upsilon\Delta}\right)
\cos\left(\lfrac{\Delta}{2}\chi-\theta\right)
-x_o\left(\lfrac{e^{\chi/2}}{\Delta}\right) 
\left[\cos\left(\lfrac{\Delta}{2}\chi-\theta\right)
      -\Delta\sin\left(\lfrac{\Delta}{2}\chi-\theta\right)\right].
\label{tmy}
\end{eqnarray}


{\bf A case of $TQ$ Squeezed States:} 
Finally, for the specific system $\{TQ,\Upsilon^2<4\omega^2,+\}$, 
the squeezed-state wave functions are obtained from 
Eq. (\cite{TMI}-103) and the functions in Table B-3.  They are
\begin{eqnarray}
{\Phi}_{\alpha,z}(x,t) & = & 
\left(\frac{1}{2\pi {Q}_1}\right)^{\frac{1}{4}}
\left(\frac
{e^{-i\frac{\Delta}{2}\chi}
   +e^{i(\frac{\Delta}{2}\chi-\theta)}\tanh{r}}
{e^{i\frac{\Delta}{2}\chi}
   +e^{-i(\frac{\Delta}{2}\chi+\theta)}\tanh{r}}
\right)^{\frac{1}{4}}
\exp\left\{-\frac{1}{4}\frac{1}{{Q}_1}
\left[x-{X}_1^+(\alpha,z)\right]^2\right\},
\nonumber\\
 &   & 
\times\exp
\left\{i\left[\frac{1}{4}\frac{{R}_1}{{Q}_1}x^2
+\frac{1}{2{Q}_1}
\left(x-\lfrac{1}{2}{X}_1^+(\alpha,z)\right){X}_1^-(\alpha,z)
\right]\right\},
\label{toss}  \\
{Q}_1 & = & \frac{1}{\Upsilon\Delta}
\left(\cosh{2r}+\cos{(\Delta\ln{\tau}-\theta)}\sinh{2r}\right),
\label{ssq1} \\
\frac{{R}_1}{{Q}_1} & = & {\Upsilon}
\frac{\cosh 2r 
+[cos(\Delta\chi -\theta)-\Delta\sin(\Delta\chi -\theta)]\sinh 2r}
{\cosh{2r}+\cos{(\Delta\chi-\theta)}\sinh{2r}},
\end{eqnarray}
where ${X}_1^+(\alpha)$ and ${X}_1^-(\alpha)$ 
are given in Eqs. (\ref{newX2})  and (\ref{newX3}). Also, 
from Eqs. (\cite{TMI}-107) and (\cite{TMI}-108), one has 
\begin{eqnarray}
{X}_1^{-}(\alpha,z) &=&
 {X}_1^{-}(\alpha) \cosh{2r} +{Y}_1^-(\alpha,\theta)\sinh{2r},
\label{ssx1-}  \\
{Y}_1^-(\alpha,\theta)  &=& 
p_o\left(\lfrac{2}{\Upsilon\Delta}\right)
\cos\left(\lfrac{\Delta}{2}\chi-\theta\right)
-x_o\left(\lfrac{1}{\Delta}\right) 
\left[\cos\left(\lfrac{\Delta}{2}\chi-\theta\right)
      -\Delta\sin\left(\lfrac{\Delta}{2}\chi-\theta\right)\right].
\label{tqy}
\end{eqnarray}


\section{Expectation Values}

\subsection{The Dynamical  Variables $\langle x \rangle$ and 
$\langle p \rangle$}

Expectation values of $\langle x \rangle$ and $\langle p \rangle$
were obtained for general time-dependent quadratic Hamiltonians 
in Ref. \cite{paperII} using algebraic methods.  
Applying these results to our specific Hamiltonians 
(see Eqs. (88), (87) and (86) of \cite{paperII}, respectively), 
we now give the 
expectation values 
$\langle x\rangle$ and $\langle p\rangle$ as a 
function of time, $\Upsilon$, and $\omega$, for the $TO$, 
$TM$, and $TQ$  systems, respectively.   
These are identical for both the coherent-state and 
squeezed-state expectation values, as they should be.  First, 
in Table 1, we give $\langle x\rangle$.

In Figures 1-3, we show examples of the quantum motion, 
$\langle x\rangle$, for the three cases:  $\Upsilon^2 ~ (>,=,<) 
~ 4\omega^2$. Each figure contains the $TO$, $TM$, and $TQ$ solutions. 

Beginning with Figure 1, we consider the case 
$\Upsilon^2 ~ > ~ 4\omega^2$.  
The $TO$ system, starting at time $(t'-t_o')=0$, 
begins at $\langle x\rangle=1$  
but then goes negative after $(t'-to')=3$.  
The $TM$ system has $\langle x\rangle$ positive but exponentially small 
for large negative time.  It reaches a maximum near 
$(t-t_o)=0$. 
Then it becomes negative, growing exponentially, for positive time.  
In the $TQ$ system, $\langle x \rangle$ is exponentially large and 
positive for large negative time, goes through unity, and is exponentially
large but negative for large positive time.  


\begin{tabular}{r|c}
\multicolumn{2}{l} {Table 1.  $\langle x\rangle$, as a 
function of $\Upsilon$ and $\omega$, for the $TO$, $TM$, and }\\
\multicolumn{2}{l} {$TQ$ systems.  $\Delta^2=|1-4\omega^2/
\Upsilon^2|$, $\tau=\left[1+\Upsilon(t'-t_o')\right]$, 
$\chi=\Upsilon(t-t_o)$.}\\*[2mm]\hline
 Class & $\langle x\rangle$ \\*[1mm]\hline
      & \\
$\Upsilon^2>4\omega^2$ &\\ 
$TO$ & $\lfrac{x_o}{\Delta}\sqrt{\tau}\left[\Delta
\cosh{\left(\lfrac{\Delta}{2}\scsz{\ln{\tau}}\right)}
-\sinh{\left(\lfrac{\Delta}{2}\scsz{\ln{\tau}}\right)}\right]
+\lfrac{2p_o}{\Upsilon\Delta}\sqrt{\tau}
\sinh{\left(\lfrac{\Delta}{2}\scsz{\ln{\tau}}\right)}$\\*[2mm]
$TM$ & $\lfrac{x_o}{\Delta}e^{\scsz{\chi/2}}\left[\Delta
\cosh{\left(\lfrac{\Delta}{2}\scsz{\chi}\right)}
-\sinh{\left(\lfrac{\Delta}{2}\scsz{\chi}\right)}\right]
+\lfrac{2p_o}{\Upsilon\Delta}e^{\scsz{\chi/2}}
\sinh{\left(\lfrac{\Delta}{2}\scsz{\chi}\right)}$ \\*[2mm]
$TQ$ & $\lfrac{x_o}{\Delta}\left[\Delta
\cosh{\left(\lfrac{\Delta}{2}\scsz{\chi}\right)}
-\sinh{\left(\lfrac{\Delta}{2}\scsz{\chi}\right)}\right]
+\lfrac{2p_o}{\Upsilon\Delta}\sinh{\left(\lfrac{\Delta}{2}\scsz{\chi}
\right)}$ \\*[2mm]\hline
  &  \\
$\Upsilon^2=4\omega^2$ & \\
 $TO$ & $x_o\sqrt{\tau}\left(1-\lfrac{1}{2}\ln{\tau}\right)
+\lfrac{p_o}{\Upsilon}\sqrt{\tau}\ln{\tau}$ \\*[2mm]
 $TM$ & $x_oe^{\scsz{\chi/2}}\left(1-\lfrac{1}{2}\chi\right)
+\lfrac{p_o}{\Upsilon}e^{\scsz{\chi/2}}\chi$ \\*[2mm]
 $TQ$ & $x_o\left(1-\lfrac{1}{2}\chi)
\right)+\lfrac{p_o}{\Upsilon}\chi$\\*[2mm]\hline
 &  \\
$\Upsilon^2 < 4\omega^2$ & \\
 $TO$ & $\lfrac{x_o}{\Delta}\sqrt{\tau}\left[\Delta
\cos{\left(\lfrac{\Delta}{2}\scsz{\ln{\tau}}\right)}
-\sin{\left(\lfrac{\Delta}{2}\scsz{\ln{\tau}}\right)}\right]
+\lfrac{2p_o}{\Upsilon\Delta}\sqrt{\tau}
\sin{\left(\lfrac{\Delta}{2}\scsz{\ln{\tau}}\right)}$\\*[2mm]
$TM$ & $\lfrac{x_o}{\Delta}e^{\scsz{\chi/2}}\left[\Delta
\cos{\left(\lfrac{\Delta}{2}\scsz{\chi}\right)}
-\sin{\left(\lfrac{\Delta}{2}\scsz{\chi}\right)}\right]
+\lfrac{2p_o}{\Upsilon\Delta}e^{\scsz{\chi/2}}
\sin{\left(\lfrac{\Delta}{2}\scsz{\chi}\right)}$ \\*[2mm]
 $TQ$ & $\lfrac{x_o}{\Delta}\left[\Delta
\cos{\left(\lfrac{\Delta}{2}\scsz{\chi}\right)}
-\sin{\left(\lfrac{\Delta}{2}\scsz{\chi}\right)}\right]
+\lfrac{2p_o}{\Upsilon\Delta}\sin{\left(\lfrac{\Delta}{2}\scsz{\chi}
\right)}$\\*[2mm]\hline
\end{tabular} 



\begin{figure}[p]
 \begin{center}
\noindent    
\psfig{figure=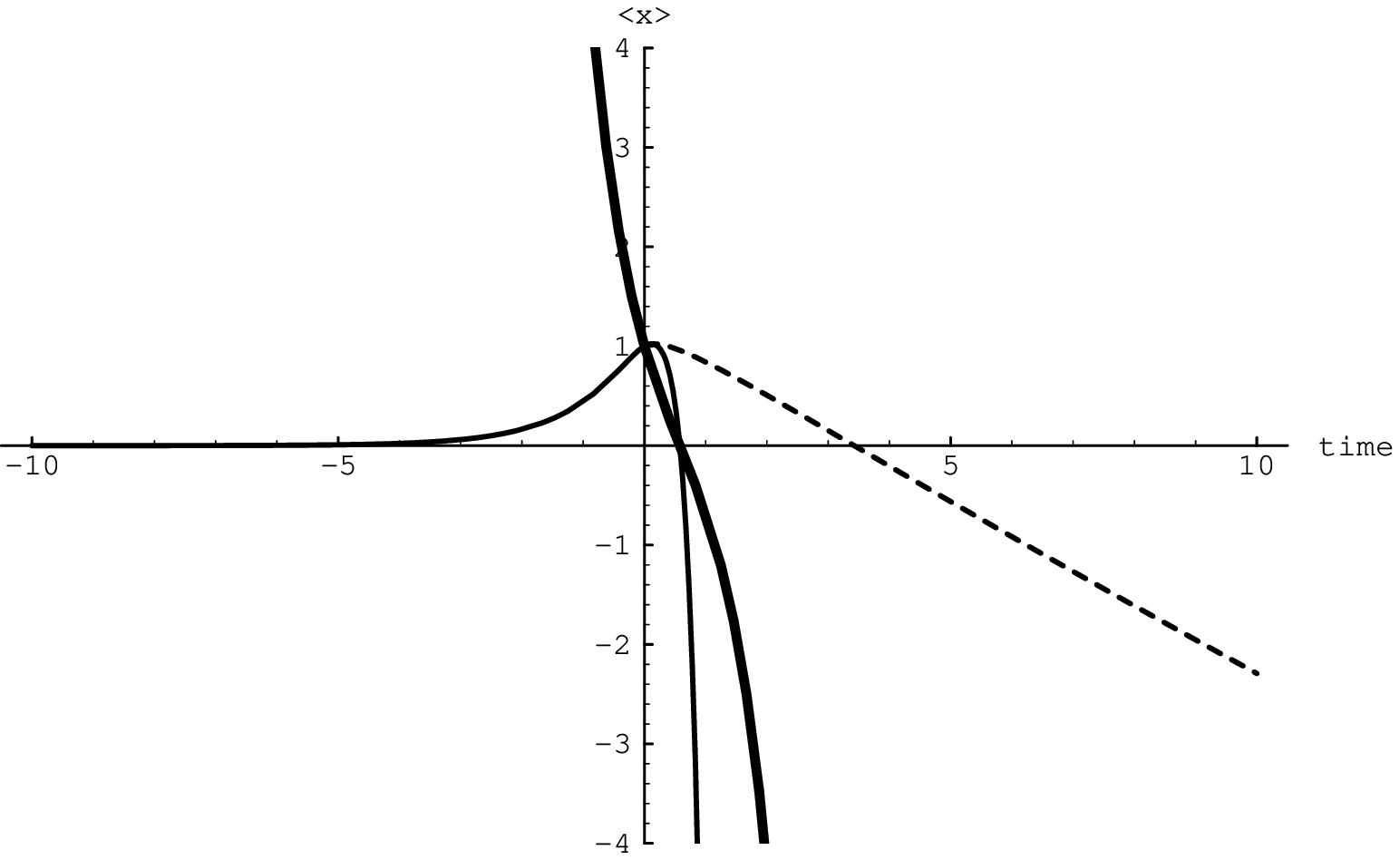,width=4in,height=3in}
  \caption{
For the case $\Upsilon^2 ~ > ~ 4\omega^2$, we show
the expectation values $\langle x\rangle$ as a function of 
time.  Time is  $(t'-t_o')$ for the $TO$ system (dashes) and $(t-t_o)$ 
for the $TM$ (line) and the $TQ$ (thick line) systems.  Here, 
$x_o = p_o =1$, $\Upsilon = 5$, and $\omega = 2$. 
The $TO$ system is restricted to $(t'-t_o')>0$.  But  
for $(t'-t_o')<0$ this curve matches on to the $TO$ curve for  
$\Upsilon < 0$; i.e.,  $\Upsilon = -5$ and $\omega = 2$.
 \label{fig:tOMQ1}}
 \end{center}
\end{figure} 



\begin{figure}[p]
 \begin{center}
\noindent    
\psfig{figure=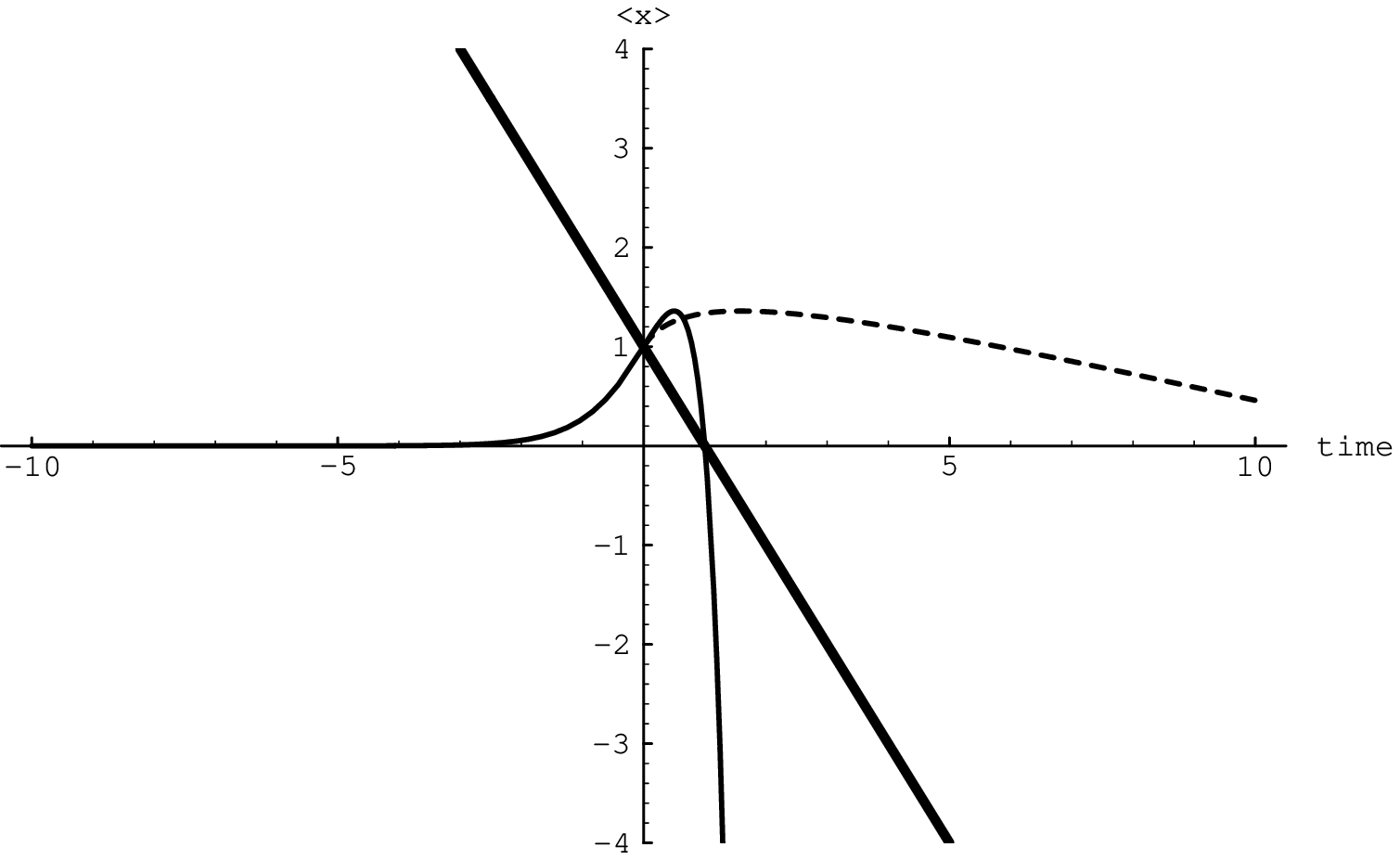,width=4in,height=3in}
  \caption{
For the case $\Upsilon^2 ~ = ~ 4\omega^2$, 
the expectation values $\langle x\rangle$ as a function of 
time.  Time is  $(t'-t_o')$ for the $TO$ system (dashes) and $(t-t_o)$ 
for the $TM$ (line) and the $TQ$ (thick line) systems.  Here, 
$x_o = p_o =1$, $\Upsilon = 4$, and $\omega = 2$. 
The $TO$ system is restricted to $(t'-t_o')>0$.  But 
for $(t'-t_o')<0$ this curve matches on to the $TO$ curve for  
$\Upsilon < 0$; i.e.,  $\Upsilon = -5$ and $\omega = 2$.
 \label{fig:tOMQ2}}

 \end{center}
\end{figure} 



\begin{figure}[ht]
 \begin{center}
\noindent    
\psfig{figure=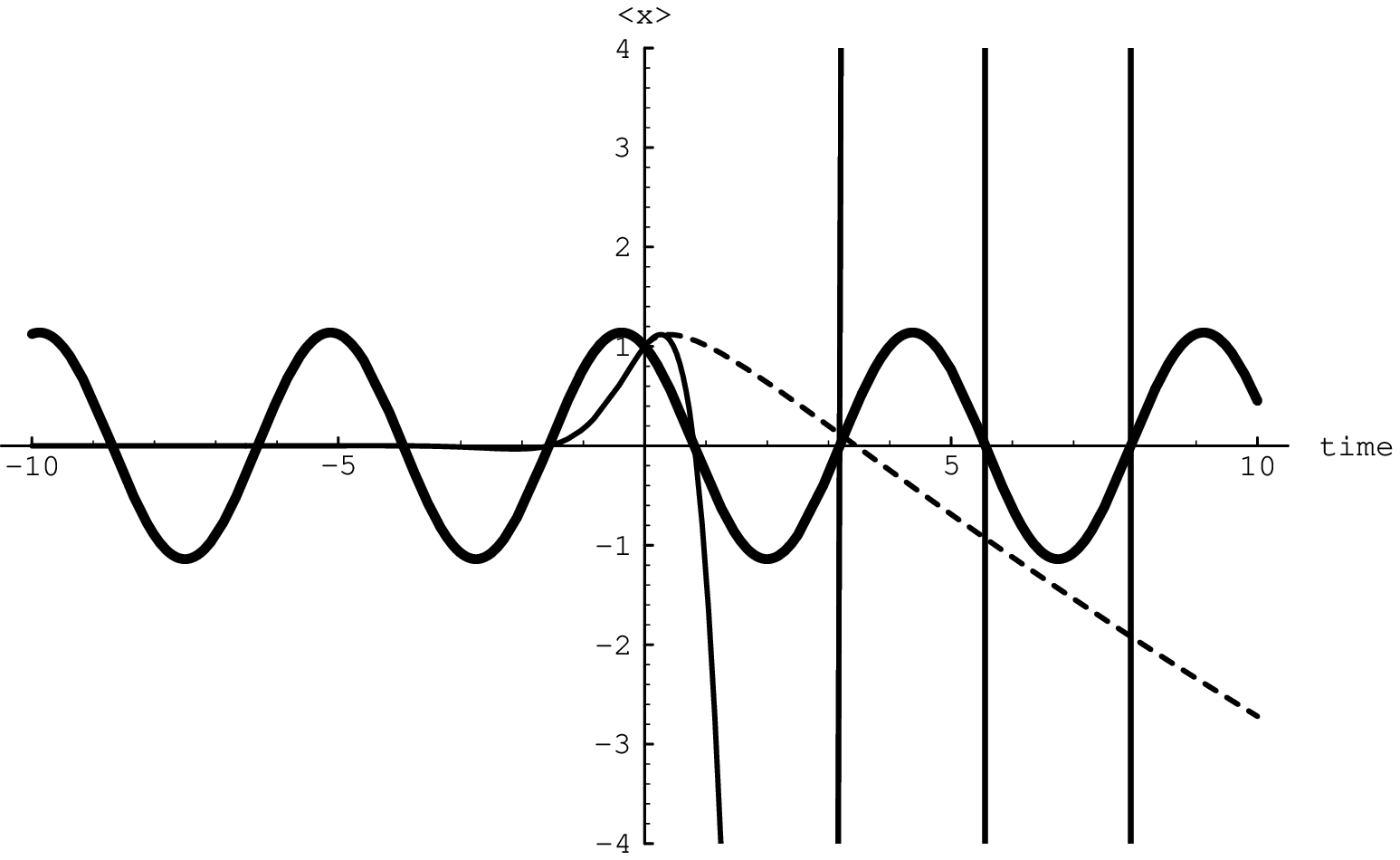,width=4in,height=3in}
  \caption{
For the case $\Upsilon^2 ~ < ~ 4\omega^2$, 
the expectation values $\langle x\rangle$ as a function of 
time.  Time is  $(t'-t_o')$ for the $TO$ system (dashes) and $(t-t_o)$ 
for the $TM$ (line) and the $TQ$ (thick line) systems.  Here, 
$x_o = p_o =1$, $\Upsilon = 3$, and $\omega = 2$. 
The $TO$ system is restricted to $(t'-t_o')>0$.  But 
for $(t'-t_o')<0$ this curve matches on to the $TO$ curve for  
$\Upsilon < 0$; i.e.,  $\Upsilon = -5$ and $\omega = 2$.
 \label{fig:tOMQ3}}
 \end{center}
\end{figure} 


Continuing to Figure 2, we have $\Upsilon^2 ~ = ~ 4\omega^2$.  We find that
the $TO$ system, starting at time $(t-t_o)=0$, 
begins at $\langle x \rangle=1$, 
rises slightly, and then decreases. Eventually 
$\langle x \rangle=1$ goes negative, near $(t'-t_o')=15$, as 
$-\sqrt{t'}\ln{t'}$.  
The $TM$ system is exponentially small 
and positive for large negative time, reaches a maximum near 
$(t-t_o)=0$, and 
then becomes exponentially large and negative for positive time.  
In the $TQ$ system, $\langle x \rangle$ varies as a straight line 
having negative slope with respect to time.  

Lastly, in Figure 3, we have $\Upsilon^2 ~ < ~ 4\omega^2$.  
The $TO$ system, starting at time $(t'-t_o')=0$, 
begins at $\langle x \rangle=1$, 
rises slightly, and then decreases,  going negative after 
$(t'-t_o')=3$.  Observe that all three $TO$ curves, up to this point, have 
been  similar in nature.  But this time there is a difference.  This curve   
has $\cos-\sin$ oscillations with respect to $\ln{t'}$. 
The curve reaches a minimum near 
$(\langle x \rangle,~(t'-t_o')) \approx (-40,~1000)$,
then reaches a maximum near $(1400,~10^6)$, and so on.  
The $TM$ system has $\langle x \rangle$ being exponentially small 
for large negative time, going through unity, and then 
turning rapidly negative.  Once again, up to this point, all three $TM$ 
curves have apparently been similar in nature.  But this curve has   $\cos-\sin$ oscillations about it.   
Therefore,  for large positive time, the exponential growth is oscillatory.  
For example, the first minimum past zero is at
$(\langle x \rangle,~(t-t_o)) \approx (-40,~2)$ 
The $TQ$ system shows normal-looking
oscillatory motion for $\langle x \rangle$. 

In Table 2 we give $\langle p \rangle$ for this problem.

\subsection{The Classical Motion}

For coherent states and squeezed states, $\langle x \rangle$ and 
$\langle p \rangle$ should obey the classical Hamiltonian 
equations of motion:
\begin{equation}
\dot{x} = \frac{\partial H}{\partial p}, ~~~~~~~~~~
\dot{p} =-\frac{\partial H}{\partial x}.   \label{ceqm}
\end{equation}

Consider the classical Hamiltonians 
associated with the Schr\"odinger equations:
\begin{eqnarray}
TO:~~~~~H&=& \frac{p^2}{2} +\frac{\omega^2}{\Upsilon^2}\frac{x^2}{2},
 \\*[1.5mm]
TM:~~~~~\hat{H}&=& e^{\Upsilon(t-t_o)}\frac{p^2}{2} 
                    +{\omega^2}e^{-\Upsilon(t-t_o)}\frac{x^2}{2},
\\*[1.5mm]
TQ:~~~~~H&=&\frac{p^2}{2} -\frac{\Upsilon}{2}xp
                 +{\omega^2}\frac{x^2}{2},
\end{eqnarray}
Applying these to the classical equations of motion (\ref{ceqm})  
one obtains (``dot" is $d/dt'$ or $d/dt$ as appropriate)
\begin{eqnarray}
&TO:&~~~~~~~\dot{ x }= p , ~~~~~~~~~~~~~~~~~~
\dot{ p } = -\lfrac{\omega^2}{\tau^2} x ,
 \\*[1.5mm]
&TM:&~~~~~~~\dot{x }=e^{\chi} p , ~~~~~~~~~~~~~~
\dot{ p } = -\omega^2 e^{-\chi} x ,
\\*[1.5mm]
&TQ:&~~~~~~~\dot{x}= p 
                 -\lfrac{\Upsilon}{2} x ,~~~~~~~~
\dot{p} = -\omega^2  x 
                 +\lfrac{\Upsilon}{2} p .
\end{eqnarray}


\begin{tabular}{r|c}
\multicolumn{2}{l} {Table 2.  $\langle p\rangle$, as a 
function of $\Upsilon$ and $\omega$, for the $TO$, $TM$, and }\\
\multicolumn{2}{l} {$TQ$ systems.  $\Delta^2=|1-4\omega^2/
\Upsilon^2|$, $\tau=\left[1+\Upsilon(t'-t_o')\right]$, 
$\chi=\Upsilon(t-t_o)$.}\\*[2mm]\hline
 Class & $\langle p\rangle$ \\*[1mm]\hline
      & \\
$\Upsilon^2>4\omega^2$ &\\
$TO$ & $-\lfrac{2x_o\omega^2}{\Upsilon\Delta}\frac{1}{\sqrt{\tau}}
\sinh{\left(\lfrac{\Delta}{2}\scsz{\ln{\tau}}\right)}+\lfrac{p_o}{\Delta}
\frac{1}{\sqrt{\tau}}\left[\Delta\cosh{\left(\lfrac{\Delta}{2}
\scsz{\ln{\tau}}\right)}
+\sinh{\left(\lfrac{\Delta}{2}\scsz{\ln{\tau}}\right)}\right]$\\*[2mm]
$TM$ & $-\lfrac{2x_o\omega^2}{\Upsilon\Delta}e^{\scsz{-\chi/2}}
\sinh{\left(\lfrac{\Delta}{2}\scsz{\chi}\right)}+\lfrac{p_o}{\Delta}
e^{\scsz{-\chi/2}}\left[\Delta\cosh{\left(\lfrac{\Delta}{2}\scsz{\chi}\right)}
+\sinh{\left(\lfrac{\Delta}{2}\scsz{\chi}\right)}\right]$ \\*[2mm]
$TQ$ & $-\lfrac{2x_o\omega^2}{\Upsilon\Delta}
\sinh{\left(\lfrac{\Delta}{2}\scsz{\chi}\right)}+\lfrac{p_o}{\Delta}
\left[\Delta\cosh{\left(\lfrac{\Delta}{2}\scsz{\chi}\right)}
+\sinh{\left(\lfrac{\Delta}{2}\scsz{\chi}\right)}\right]$ \\*[2mm]\hline
  &  \\
$\Upsilon^2=4\omega^2$ & \\
 $TO$ & $-x_o\lfrac{\Upsilon}{4}\frac{1}{\sqrt{\tau}}\ln{\tau}
+p_o\frac{1}{\sqrt{\tau}}\left(1+\lfrac{1}{2}\ln{\tau}\right)$ \\*[2mm]
 $TM$ & $-x_o\lfrac{\Upsilon}{4}e^{\scsz{-\chi/2}}\chi+p_oe^{\scsz{-\chi/2}}
\left(1+\lfrac{1}{2}\chi\right)$ \\*[2mm]
 $TQ$ & $-x_o\lfrac{\Upsilon}{4}\chi+p_o\left(1+\lfrac{1}{2}\chi)
\right)$\\*[2mm]\hline
 &  \\
$\Upsilon^2 < 4\omega^2$ & \\
 $TO$ & $-\lfrac{2x_o\omega^2}{\Upsilon\Delta}\frac{1}{\sqrt{\tau}}
\sin{\left(\lfrac{\Delta}{2}\scsz{\ln{\tau}}\right)}+\lfrac{p_o}{\Delta}
\lfrac{1}{\sqrt{\tau}}\left[\Delta\cos{\left(\lfrac{\Delta}{2}
\scsz{\ln{\tau}}\right)}+\sin{\left(\lfrac{\Delta}{2}\scsz{\ln{\tau}}
\right)}\right]$\\*[2mm]
 $TM$ & $-\lfrac{2x_o\omega^2}{\Upsilon\Delta}e^{\scsz{-\chi/2}}
\sin{\left(\lfrac{\Delta}{2}\scsz{\chi}\right)}+\lfrac{p_o}{\Delta}
e^{\scsz{-\chi/2}}\left[\Delta\cos{\left(\lfrac{\Delta}{2}\scsz{\chi}
\right)}+\sin{\left(\lfrac{\Delta}{2}\scsz{\chi}\right)}\right]$ \\*[2mm]
 $TQ$ & $-\lfrac{2x_o\omega^2}{\Upsilon\Delta}
\sin{\left(\lfrac{\Delta}{2}\scsz{\chi}\right)}+\lfrac{p_o}{\Delta}
\left[\Delta\cos{\left(\lfrac{\Delta}{2}\scsz{\chi}\right)}
+\sin{\left(\lfrac{\Delta}{2}\scsz{\chi}\right)}\right]$\\*[2mm]\hline
\end{tabular}


The reader can verify that all the expectation values in Tables 1 and 
2 satisfy these equations of motion.  This specifically demonstrates 
the general results for quadratic time-dependent Hamiltonians derived in 
\cite{paperII}.


\subsection{Uncertainties}

Similarly, one can calculate $\langle x^2 \rangle$ and $\langle p^2 \rangle$ 
and hence the uncertainties $(\Delta x)^2$ and $(\Delta p)^2$.
The uncertainty products $(\Delta x)^2$$(\Delta p)^2$ for 
all three types of Schr\"odinger equations then follow.  These are shown in 
Tables 3, 4, and 5.


\begin{tabular}{r|c}
\multicolumn{2}{l} {Table 3.  $(\Delta x)^2$ for the $TO$, 
$TM$, and $TQ$ systems when $\Upsilon > 0$.}\\
\multicolumn{2}{l} {For $\Upsilon < 0$, change $\Upsilon$ 
to $|\Upsilon|$ and $\theta$ to $-\theta$ in the following.
}\\
\multicolumn{2}{l} {$\Delta^2=|1-4\omega^2/
\Upsilon^2|$, $\tau=\left[1+\Upsilon(t'-t_o')\right]$, 
$\chi=\Upsilon(t-t_o)$, $s=\exp{r}$.}\\*[2mm]\hline
 Class & $(\Delta x)^2$ \\*[1mm]\hline
      & \\
$\Upsilon^2>4\omega^2$ &\\
$TO$ & $\frac{\tau}{2\Upsilon\Delta}\left\{s^2\left[
\cosh{(\Delta\ln{\tau})}-\sinh{(\Delta\ln{\tau})}\cos{\theta}+
\sin{\theta}\right]\right.$\\
  & $\left.\hspace{1cm}+\frac{1}{s^2}\left[\cosh{(\Delta\ln{\tau})}+
\sinh{(\Delta\ln{\tau})}\cos{\theta}-\sin{\theta}\right]\right\}$\\*[2mm]
$TM$ & $\frac{e^{\chi}}{2\Upsilon\Delta}\left\{s^2\left[\cosh{(\Delta\chi)}-
\sinh{(\Delta\chi)}\cos{\theta}+\sin{\theta}\right]\right.$\\
     & $\hspace{1cm}\left.+\frac{1}{s^2}\left[\cosh{(\Delta\chi)}+
\sinh{(\Delta\chi)}\cos{\theta}-\sin{\theta}\right]\right\}$\\*[2mm]
$TQ$ & $\frac{1}{2\Upsilon\Delta}\left\{s^2\left[\cosh{(\Delta\chi)}-
\sinh{(\Delta\chi)}\cos{\theta}+\sin{\theta}\right]\right.$\\
     & $\hspace{1cm}\left.+\frac{1}{s^2}\left[\cosh{(\Delta\chi)}+
\sinh{(\Delta\chi)}\cos{\theta}-\sin{\theta}\right]\right\}$\\*[2mm]\hline
  &  \\
$\Upsilon^2=4\omega^2$ & \\
 $TO$ & $\frac{\tau}{4\Upsilon}\left\{s^2\left[(1+\ln^2{\tau})+
(1-\ln^2{\tau})\cos{\theta}+2\ln{\tau}\sin{\theta}\right]\right.$\\
      & $\hspace{1cm}\left.+\frac{1}{s^2}\left[(1+\ln^2{\tau})-
(1-\ln^2{\tau})\cos{\theta}-2\ln{\tau}\sin{\theta}\right]\right\}$\\*[2mm]
 $TM$ & $\frac{e^{\chi}}{4\Upsilon}\left\{s^2\left[(1+\chi^2)+(1-\chi^2)
\cos{\theta}+2\chi\sin{\theta}\right]\right.$\\
      & $\hspace{1cm}\left.+\frac{1}{s^2}\left[(1+\chi^2)-(1-\chi^2)
\cos{\theta}-2\chi\sin{\theta}\right]\right\}$\\*[2mm]
 $TQ$ & $\frac{1}{4\Upsilon}\left\{s^2\left[(1+\chi^2)+(1-\chi^2)
\cos{\theta}+2\chi\sin{\theta}\right]\right.$\\
      & $\left.+\frac{1}{s^2}\left[(1+\chi^2)-(1-\chi^2)
\cos{\theta}-2\chi\sin{\theta}\right]\right\}$\\*[2mm]\hline
 &  \\
$\Upsilon^2 < 4\omega^2$ & \\
 $TO$ & $\frac{\tau}{2\Upsilon\Delta}\left\{s^2\left[1+\cos{(\Delta
\ln{\tau}-\theta)}\right]+\frac{1}{s^2}\left[1-\cos{(\Delta\ln{\tau}
-\theta)}\right]\right\}$\\*[2mm]
 $TM$ & $\frac{e^{\chi}}{2\Upsilon\Delta}\left\{s^2\left[1+\cos{(\Delta
\chi-\theta)}\right]+\frac{1}{s^2}\left[1-\cos{(\Delta\chi
-\theta)}\right]\right\}$\\*[2mm]
 $TQ$ & $\frac{1}{2\Upsilon\Delta}\left\{s^2\left[1+\cos{(\Delta
\chi-\theta)}\right]+\frac{1}{s^2}\left[1-\cos{(\Delta\chi
-\theta)}\right]\right\}$\\*[2mm]\hline
\end{tabular} 

\begin{tabular}{r|ll}
\multicolumn{3}{l} {Table 4.  $(\Delta p)^2$ for the $TO$, 
$TM$, and $TQ$ systems when $\Upsilon > 0$.}\\
\multicolumn{3}{l} {For $\Upsilon < 0$, change $\Upsilon$ 
to $|\Upsilon|$ and $\theta$ to $-\theta$ in the following.
}\\
\multicolumn{3}{l} {$\Delta^2=|1-4\omega^2/
\Upsilon^2|$, $\tau=\left[1+\Upsilon(t'-t_o')\right]$, 
$\chi=\Upsilon(t-t_o)$, $s=\exp{r}$.}\\*[2mm]\hline
 Class &  & \multicolumn{1}{c} {$(\Delta p)^2$} \\*[1mm]\hline
      &  &\\
$\Upsilon^2>4\omega^2$ & &\\
$TO$ & \hspace{9mm}  & \scsz{\frac{\Upsilon}{8\Delta\tau}\left\{s^2\left\{
(1+\Delta^2)\left[\cosh{(\Delta\ln{\tau})}-\sinh{(\Delta\ln{\tau})}
\cos{\theta}\right]\right.\right.}\\
     & & \scsz{\left.\left.\hspace{1cm}+(1-\Delta^2)\sin{\theta}+2\Delta
\left[\sinh{(\Delta\ln{\tau})}-
\cosh{(\Delta\ln{\tau})}\cos{\theta}\right]\right\}\right.}\\
     & & \scsz{\left.\hspace{1cm}+\frac{1}{s^2}\left\{(1+\Delta^2)\left[
\cosh{(\Delta\ln{\tau})}+\sinh{(\Delta\ln{\tau})}\cos{\theta}\right]
\right.\right.}\\
     & & \scsz{\left.\left.\hspace{1cm}-(1-\Delta^2)\sin{\theta}+2\Delta
\left[\sinh{(\Delta\ln{\tau})}
+\cosh{(\Delta\ln{\tau})}\cos{\theta}\right]\right\}\right\}}\\*[2mm]
$TM$ & & \scsz{\frac{\Upsilon e^{-\chi}}{8\Delta}\left\{s^2\left\{(1+\Delta^2)
\left[\cosh{(\Delta\chi)}-\sinh{(\Delta\chi)}\cos{\theta}\right]
\right.\right.}\\
     & & \scsz{\hspace{1cm}\left.\left.+(1-\Delta^2)\sin{\theta}+2\Delta
\left[\sinh{(\Delta\chi)}-\cosh{(\Delta\chi)}\cos{\theta}\right]\right\}
\right.}\\
     & & \scsz{\left.\hspace{1cm}+\frac{1}{s^2}\left\{(1+\Delta^2)\left[
\cosh{(\Delta\chi)}+\sinh{(\Delta\chi)}\cos{\theta}\right]\right.
\right.}\\
     & & \scsz{\left.\left.\hspace{1cm}-(1-\Delta^2)\sin{\theta}
+2\Delta\left[\sinh{(\Delta\chi)}+\cosh{(\Delta\chi)}\cos{\theta}\right]
\right\}\right\}}\\*[2mm]
$TQ$ & & \scsz{\frac{\Upsilon}{8\Delta}\left\{s^2\left\{(1+\Delta^2)\left[
\cosh{(\Delta\chi)}-\sinh{(\Delta\chi)}\cos{\theta}\right]\right.
\right.}\\
     & & \scsz{\left.\left.+(1-\Delta^2)\sin{\theta}+2\Delta\left[
\sinh{(\Delta\chi)}-\cosh{(\Delta\chi)}\cos{\theta}\right]\right\}\right.}\\
     & & \scsz{\left.\hspace{1cm}+\frac{1}{s^2}\left\{(1+\Delta^2)\left[
\cosh{(\Delta\chi)}+\sinh{(\Delta\chi)}\cos{\theta}\right]\right.
\right.}\\
     & & \scsz{\left.\left.\hspace{1cm}-(1-\Delta^2)\sin{\theta}+2\Delta\left[
\sinh{(\Delta\chi)}+\cosh{(\Delta\chi)}\cos{\theta}\right]\right\}
\right\}}\\*[2mm]\hline
  &  &\\
$\Upsilon^2=4\omega^2$ & &\\
 $TO$ & & \scsz{\frac{\Upsilon}{4\tau}\left\{s^2\left\{\left[\frac{1}{4}+
(1+\frac{1}{2}\ln{\tau})^2\right]+
\left[\frac{1}{4}-(1+\frac{1}{2}\ln{\tau})^2\right]\cos{\theta}
+(1+\lfrac{1}{2}\ln{\tau})\sin{\theta}\right\}\right.}\\
      & & \scsz{\hspace{1cm}\left.+\frac{1}{s^2}\left\{\left[\frac{1}{4}+(1+
\frac{1}{2}\ln{\tau})^2\right]-\left[\frac{1}{4}-(1+\frac{1}{2}
\ln{\tau})^2\right]\cos{\theta}-(1+\frac{1}{2}\ln{\tau})
\sin{\theta}\right\}\right\}}\\*[2mm]
 $TM$ & & \scsz{\frac{\Upsilon e^{-\chi}}{4}\left\{s^2\left\{\left[\frac{1}{4}+
(1+\lfrac{1}{2}\chi)^2\right]+\left[\frac{1}{4}-(1-\frac{1}{2}\chi)^2\right]
\cos{\theta}+\left(1+\frac{1}{2}\chi\right)\sin{\theta}
\right\}\right.}\\
      & & \scsz{\hspace{1cm}\left.+\frac{1}{s^2}\left\{\left[\frac{1}{4}+(1+
\frac{1}{2}\chi)^2\right]-\left[\frac{1}{4}-(1+\frac{1}{2}\chi)^2\right]
\cos{\theta}-(1+\frac{1}{2}\chi)\sin{\theta}
\right\}\right\}}\\*[2mm]
 $TQ$ & & \scsz{\frac{\Upsilon}{4}\left\{s^2\left\{\left[\frac{1}{4}+(1+
\frac{1}{2}\chi)^2\right]+\left[\frac{1}{4}-(1+\frac{1}{2}\chi)^2\right]
\cos{\theta}+(1+\frac{1}{2}\chi)\sin{\theta}
\right\}\right.}\\
      & & \scsz{\hspace{1cm}\left.+\frac{1}{s^2}\left\{\left[\frac{1}{4}+(1+
\frac{1}{2}\chi)^2\right]-\left[\frac{1}{4}-(1+\frac{1}{2}\chi)^2\right]
\cos{\theta}-(1+\frac{1}{2}\chi)\sin{\theta}\right\}
\right\}}\\*[2mm]\hline
 &   & \\
$\Upsilon^2 < 4\omega^2$ & &\\
 $TO$ & & \scsz{\frac{\Upsilon}{8\Delta\tau}\left\{s^2\left[(1+\Delta^2)
+(1-\Delta^2)\cos{(\Delta\ln{\tau}-\theta)}-2\Delta\sin{(\Delta\ln{\tau}
-\theta)}\right]\right.}\\
      & & \scsz{\left.\hspace{1cm}+\frac{1}{s^2}\left[(1+\Delta^2)-(1-\Delta^2)
\cos{(\Delta\ln{\tau}-\theta)}+2\Delta\sin{(\Delta\ln{\tau}-\theta)}
\right]\right\}}\\*[2mm]
 $TM$ & & \scsz{\frac{\Upsilon e^{-\chi}}{8\Delta}\left\{s^2\left[(1+\Delta^2)+
(1-\Delta^2)\cos{(\Delta\chi-\theta)}-2\Delta\sin{(\Delta\chi-\theta)}
\right]\right.}\\
      & & \scsz{\left.\hspace{1cm}+\frac{1}{s^2}\left[(1+\Delta^2)-
(1-\Delta^2)\cos{(\Delta\chi-\theta)}+2\Delta\sin{(\Delta\chi-\theta)}
\right]\right\}}\\*[2mm]
 $TQ$ & & \scsz{\frac{\Upsilon}{8\Delta}\left\{s^2\left[(1+\Delta^2)+
(1-\Delta^2)\cos{(\Delta\chi-\theta)}-2\Delta\sin{(\Delta\chi-\theta)}
\right]\right.}\\
      & & \scsz{\left.+\frac{1}{s^2}\left[(1+\Delta^2)-(1-\Delta^2)
\cos{(\Delta\chi-\theta)}+2\Delta\sin{(\Delta\chi-\theta)}\right]
\right\}}\\*[2mm]\hline
\end{tabular} 



\begin{tabular}{r|ll}
\multicolumn{3}{l} {Table 5.  $(\Delta x)^2(\Delta p)^2$ for 
the $TO$, $TM$, and $TQ$ systems}\\
\multicolumn{3}{l} {when $\Upsilon > 0$.  
For $\Upsilon < 0$, 
replace $\theta$ by $-\theta$  in the following.}\\
\multicolumn{3}{l} {$\Delta^2=|1-4\omega^2/
\Upsilon^2|$, $\tau=\left[1+\Upsilon(t'-t_o')\right]$, 
$\chi=\Upsilon(t-t_o)$, $s=\exp{r}$. 
}\\*[2mm]\hline
 Class &  & \multicolumn{1}{c} {$(\Delta p)^2$} \\*[1mm]\hline
      &  &\\
$\Upsilon^2>4\omega^2$ & &\\
$TO$ & \hspace{9mm}  & \scsz{\frac{1}{4}\left\{1+\frac{1}{4\Delta^2}
\left\{s^2\left[\sin{\theta}+(1-\Delta\cos{\theta})
\cosh{(\Delta\ln{\tau})}+(\Delta-\cos{\theta})\sinh{(\Delta\ln{\tau})}
\right]\right.\right.}\nonumber\\
     & & \scsz{\left.\left.\hspace{1cm}+\frac{1}{s^2}\left[-\sin{\theta}
+(1+\Delta\cos{\theta})\cosh{(\Delta\ln{\tau})}+(\Delta-\cos{\theta})
\sinh{(\Delta\ln{\tau})}\right]\right\}^2\right\}}\\*[2mm]
$TM$ & & \scsz{\frac{1}{4}\left\{1+\frac{1}{4\Delta^2}\left\{
s^2\left[\sin{\theta}+(1-\Delta\cos{\theta}\cosh{(\Delta\chi)}
+(\Delta-\cos{\theta})\sinh{(\Delta\chi)}\right]\right.\right.}\\
     & & \scsz{\hspace{1cm}\left.\left.\frac{1}{s^2}\left[-\sin{\theta}
+(1+\Delta\cos{\theta})\cosh{(\Delta\chi)}+(\Delta-\cos{\theta})
\sinh{(\Delta\chi)}\right]\right\}^2\right\}}\\*[2mm]
$TQ$ & &\scsz{\frac{1}{4}\left\{1+\frac{1}{4\Delta^2}\left\{
s^2\left[\sin{\theta}+(1-\Delta\cos{\theta}\cosh{(\Delta\chi)}
+(\Delta-\cos{\theta})\sinh{(\Delta\chi)}\right]\right.\right.}\\
     & & \scsz{\hspace{1cm}\left.\left.\frac{1}{s^2}\left[-\sin{\theta}
+(1+\Delta\cos{\theta})\cosh{(\Delta\chi)}+(\Delta-\cos{\theta})
\sinh{(\Delta\chi)}\right]\right\}^2\right\}}\\*[2mm]\hline 
  &  &\\
$\Upsilon^2=4\omega^2$ & &\\
 $TO$ & & \scsz{\frac{1}{4}\left\{1+\frac{1}{4}\left\{s^2\left[
\frac{1}{2}(1+\ln{\tau})^2+(1+\ln{\tau})\sin{\theta}+\frac{1}{2}
(1-2\ln{\tau}-\ln^2{\tau})\cos{\theta}\right]\right.\right.}\\
      & & \scsz{\hspace{1cm}\left.\left.+\frac{1}{s^2}\left[
\frac{1}{2}(1+\ln{\tau})^2-(1+\ln{\tau})\sin{\theta}
-\frac{1}{2}(1-2\ln{\tau}-\ln^2{\tau})\cos{\theta}\right]\right\}^2
\right\}}\\*[2mm]
 $TM$ & & \scsz{\frac{1}{4}\left\{1+\frac{1}{4}\left\{s^2\left[
\frac{1}{2}(1+\chi)^2+(1+\chi)\sin{\theta}+\frac{1}{2}(1-2\chi-\chi^2)
\cos{\theta}\right]\right.\right.}\\
      & & \scsz{\hspace{1cm}\left.\left.+\frac{1}{s^2}\left[
\frac{1}{2}(1+\chi)^2-(1+\chi)\sin{\theta}-\frac{1}{2}(1-2\chi-\chi^2)
\cos{\theta}\right]\right\}^2\right\}}\\*[2mm]
 $TQ$ & & \scsz{\frac{1}{4}\left\{1+\frac{1}{4}\left\{s^2\left[
\frac{1}{2}(1+\chi)^2+(1+\chi)\sin{\theta}+\frac{1}{2}(1-2\chi-\chi^2)
\cos{\theta}\right]\right.\right.}\\
      & & \scsz{\hspace{1cm}\left.\left.+\frac{1}{s^2}\left[
\frac{1}{2}(1+\chi)^2-(1+\chi)\sin{\theta}-\frac{1}{2}(1-2\chi-\chi^2)
\cos{\theta}\right]\right\}^2\right\}}\\*[2mm]\hline
 &   & \\
$\Upsilon^2 < 4\omega^2$ & &\\
 $TO$ & & \scsz{\frac{1}{4}\left\{1+\frac{1}{4\Delta^2}\left\{s^2\left[
1+\cos{(\Delta\ln{\tau}-\theta)}-\Delta\sin{(\Delta\ln{\tau}-\theta)}
\right]\right.\right.}\\
      & & \scsz{\left.\left.\hspace{1cm}+\frac{1}{s^2}\left[1-
\cos{(\Delta\ln{\tau}-\theta)}+\Delta\sin{(\Delta\ln{\tau}-\theta)}
\right]\right\}^2\right\}}\\*[2mm]
 $TM$ & & \scsz{\frac{1}{4}\left\{1+\frac{1}{4\Delta^2}\left\{s^2\left[
1+\cos{(\Delta\chi-\theta)}-\Delta\sin{(\Delta\chi-\theta)}\right]
\right.\right.}\\
      & & \scsz{\left.\left.\hspace{1cm}+\frac{1}{s^2}\left[
1-\cos{(\Delta\chi-\theta)}+\Delta\sin{(\Delta\chi-\theta)}
\right]\right\}^2\right\}}\\*[2mm]
 $TQ$ & & \scsz{\frac{1}{4}\left\{1+\frac{1}{4\Delta^2}\left\{s^2\left[
1+\cos{(\Delta\chi-\theta)}-\Delta\sin{(\Delta\chi-\theta)}\right]
\right.\right.}\\
      & & \scsz{\left.\left.\hspace{1cm}+\frac{1}{s^2}\left[
1-\cos{(\Delta\chi-\theta)}+\Delta\sin{(\Delta\chi-\theta)}
\right]\right\}^2\right\}}\\*[2mm]\hline
\end{tabular} 



\section{Conclusion}

Our results extend and complement the seminal work of Dodonov and Man'ko 
{\cite{dm1}} as well as the work of Cheng and Fung {\cite{cf1}}.  These 
authors   
focused on solving the Schr\"odinger equation with Hamiltonian (\ref{oureq}).  
In both papers the authors use a symmetry approach.  Dodonov and Man'ko 
obtained a set of number states similar to ours for the weakly damped case 
($\Upsilon^2 < 4\omega^2$), the strongly damped case 
($\Upsilon^2 > 4\omega^2$), and the criticaly damped case 
($\Upsilon^2 = 4\omega^2$).  For each damping case  
Cheng and Fung used evolution operator and Lie 
methods to compute the expectation values of $x$ and $p$ in coherent 
and squeezed states.  

In our work, we have shown how Lie symmetry methods can be used to obtain 
number states, coherent states, and squeezed states not only for the 
$TM$-type Schr\"odinger equations with Hamiltonian (\ref{oureq}), but also 
for systems of related $TQ$ and $TO$ equations.  The general transformation  
method developed in {\cite{paperI,paperII,TMI}} demonstrates the intimate 
connection between these classes of Schr\"odinger equations, their solutions, 
and the expectation values.  Furthermore, for each class of Schr\"odinger 
equation, we have proven that the coherent- and squeezed-state expectation 
values of position and momentum satisfy Hamilton's equations.


{\bf Acknowledgements:}
MMN acknowledges the support of the United States Department of 
Energy.  DRT acknowledges
a grant from the Natural Sciences and Engineering Research Council 
of Canada.


\section*{Appendix A: Solutions to Eq. (\ref{e:hu12a}) }

The second-order ordinary differential equation 
\begin{equation}
\frac{d^2w}{ds^2}+As^{-2}w=0,\label{e:ApA1}
\end{equation}
is an Euler-type equation with a regular singular point at $s=0$. 
[The Wronskian of the solutions to Eq. (\ref{e:ApA1}) is a constant 
{\cite{aec1}}.] Assuming that $s > 0$, 
we look for solutions of the form 
\begin{equation}
w(s)=s^k.\label{e:ApA4}
\end{equation}
Substituting (\ref{e:ApA4}) into Eq. (\ref{e:ApA1}) yields a 
quadratic for $k$ and  solutions 
\begin{equation}
k(k-1)+A=0, ~~~~~~~~~~~~~~~~
k=\lfrac{1}{2}\left(1\pm \sqrt{1-4A}\right).\label{e:ApA12} 
\end{equation}
The solutions are characterized by
value and sign of the discriminant $1-4A$.  
\begin{equation}
k=\lfrac{1}{2}\left(1\pm \sigma\Delta\right),~~~~~~~~~~
\sigma=\sqrt{{\rm sign}(1-4A)},~~~~
\Delta^2=|1-4A|.
\label{e:ApA20}
\end{equation}

Now, let us examine the various real solutions of Eq. (\ref{e:ApA1}). 

{\bf (i)} $1-4A > 0$: 
In this case, $\sigma=1$.  The real solutions and their Wronskian are 
\begin{eqnarray}
w_1(s) & = &  s^{(1-\Delta)/2}
        =  \sqrt{s}\exp{\left(-\lfrac{\Delta}{2}\ln{s}\right)},
 \label{e:ApA1r1.1} \\
w_2(s) & = & s^{(1+\Delta)/2}
        =  \sqrt{s}\exp{\left(\lfrac{\Delta}{2}\ln{s}\right)}.
\label{e:ApA1r1.2} \\
& & W_s(w_1,w_2)=\Delta.\label{e:ApA1rw1}
\end{eqnarray}

{\bf (ii)} $1-4A=0$:
In this case, the two roots are equal.  The two solutions are 
\begin{equation}
w_1=\sqrt{s},~~~~~~~~~~~~~~~w_2=\sqrt{s}\ln{s}.\label{e:ApA1r2.2}
\end{equation}
The second is obtained by standard methods {\cite{aec1}} 
The Wronskian of these two solutions is 
\begin{equation}
W_s(w_1,w_2)=1.\label{e:ApA1rw2.2}
\end{equation}

{\bf (iii)} $1-4A < 0$: In this case, 
$\sigma=i$ and we obtain two  solutions  
that are complex conjugates of one another.  
By taking appropriate linear combinations of the two 
complex solutions, we obtain real solutions $w_1$ and $w_2$
\begin{equation}
w_1(s)=\sqrt{s}\cos{\left(\lfrac{\Delta}{2}
\ln{s}\right)},~~~~~~~~~~w_2(s)=\sqrt{s}
\sin{\left(\lfrac{\Delta}{2}\ln{s}\right)}.\label{e:ApA1r3.1}
\end{equation}
Their Wronskian is given by 
\begin{equation}
W_s(w_1,w_2)=\Delta/2.\label{e:ApA1rw3.1}
\end{equation}

The nonzero Wronskians demonstrate  that in all cases the two real solutions
are linearly independent of each other {\cite{aec1}}.
\newpage


\section*{Appendix B: Time-dependent $TO$, $TM$, and $TQ$ functions}


\begin{tabular}{c|c|c}
\multicolumn{3}{l} {Table B-1.  Time dependent functions 
for the $TO$ system. The}\\
\multicolumn{3}{l} {functions $\bar{\xi}$, $\dot{\bar{\xi}}$, $\phi_2$, and 
$\dot{\phi}_2$ can be obtained by taking the complex conjugate of}\\
\multicolumn{3}{l} {$\xi$, $\dot{\xi}$, $\phi_1$, and $\dot{\phi}_1$, 
respectively. $\Delta^2=|1-4\omega^2/\Upsilon^2|$, 
$\tau=\left[1+\Upsilon(t'-t_o')\right]$.}\\*[2mm]\hline
 & $\Upsilon > 0$ & $\Upsilon < 0$\\*[1mm]\hline
 & \multicolumn{2}{c} {$\Upsilon^2>4\omega^2$} \\*[1mm]\cline{2-3}
\scsz{\xi(t')} & \scsz{\sqrt{\lfrac{1}{2\Upsilon\Delta}}\sqrt{\tau}
\left(e^{-\lfrac{\Delta}{2}\ln{\tau}}+ie^{\lfrac{\Delta}{2}\ln{\tau}}
\right)} & \scsz{\sqrt{\lfrac{1}{2\left|\Upsilon\right|\Delta}}\sqrt{\tau}
\left(e^{-\lfrac{\Delta}{2}\ln{\tau}}-ie^{\lfrac{\Delta}{2}\ln{\tau}}
\right)}  \\*[2mm]
\scsz{\dot{\xi}(t')} & \scsz{\sqrt{\lfrac{\Upsilon}{8\Delta}}
\frac{1}{\sqrt{\tau}}\left[(1-\Delta)e^{-\lfrac{\Delta}{2}\ln{\tau}}
+i(1+\Delta)e^{\lfrac{\Delta}{2}\ln{\tau}}\right]} & 
\scsz{-\sqrt{\lfrac{\left|\Upsilon\right|}{8\Delta}}
\frac{1}{\sqrt{\tau}}\left[(1-\Delta)e^{-\lfrac{\Delta}{2}\ln{\tau}}
-i(1+\Delta)e^{\lfrac{\Delta}{2}\ln{\tau}}\right]}\\*[2mm]
\scsz{\phi_3(t')} & \scsz{\lfrac{1}{\Upsilon\Delta}\tau\left(
e^{-\Delta\ln{\tau}}+e^{\Delta\ln{\tau}}\right)} & 
\scsz{\lfrac{1}{\left|\Upsilon\right|\Delta}\tau\left(
e^{-\Delta\ln{\tau}}+e^{\Delta\ln{\tau}}\right)}\\*[2mm]
\scsz{\dot{\phi}_3(t')} & \scsz{\lfrac{1}{\Delta}\left[(1-\Delta)
e^{-\Delta\ln{\tau}}+(1+\Delta)e^{\Delta\ln{\tau}}\right]} &  
\scsz{-\lfrac{1}{\Delta}\left[(1-\Delta)
e^{-\Delta\ln{\tau}}+(1+\Delta)e^{\Delta\ln{\tau}}\right]} \\*[2mm]
\scsz{\ddot{\phi}_3(t')} & \scsz{\frac{\Upsilon}{\tau}\left[-(1-\Delta)
e^{-\Delta\ln{\tau}}+(1+\Delta)e^{\Delta\ln{\tau}}\right]} & 
\scsz{\frac{\left|\Upsilon\right|}{\tau}\left[-(1-\Delta)
e^{-\Delta\ln{\tau}}+(1+\Delta)e^{\Delta\ln{\tau}}\right]} \\*[2mm]
\scsz{\phi_1(t')} & \scsz{\lfrac{1}{2\Upsilon\Delta}\tau\left(
e^{-\Delta\ln{\tau}}-e^{\Delta\ln{\tau}}+2i\right)} & 
\scsz{\lfrac{1}{2\left|\Upsilon\right|\Delta}\tau\left(
e^{-\Delta\ln{\tau}}-e^{\Delta\ln{\tau}}-2i\right)}\\*[2mm]
\scsz{\dot{\phi}_1(t')} & \scsz{\lfrac{1}{2\Delta}\left[(1-\Delta)
e^{-\Delta\ln{\tau}}-(1+\Delta)e^{\Delta\ln{\tau}}+2i\right]} & 
\scsz{-\lfrac{1}{2\Delta}\left[(1-\Delta)
e^{-\Delta\ln{\tau}}-(1+\Delta)e^{\Delta\ln{\tau}}-2i\right]} \\*[2mm]\hline
 & \multicolumn{2}{c} {$\Upsilon^2 = 4\omega^2$} \\*[1mm]\cline{2-3}
\scsz{\xi(t')} & \scsz{\sqrt{\lfrac{1}{2\Upsilon}}\sqrt{\tau}
\left(1+i\ln{\tau}\right)} & \scsz{\sqrt{\lfrac{1}{2\left|\Upsilon\right|}}
\sqrt{\tau}\left(1-i\ln{\tau}\right)}\\*[2mm]
\scsz{\dot{\xi}(t')} & \scsz{\sqrt{\lfrac{\Upsilon}{2}}
\frac{1}{\sqrt{\tau}}\left[\lfrac{1}{2}+i\left(1+\lfrac{1}{2}
\ln{\tau}\right)\right]} & \scsz{-\sqrt{\lfrac{\left|\Upsilon\right|}{2}}
\frac{1}{\sqrt{\tau}}\left[\lfrac{1}{2}-i\left(1+\lfrac{1}{2}
\ln{\tau}\right)\right]}\\*[2mm]
\scsz{\phi_3(t')} & \scsz{\lfrac{1}{\Upsilon}\tau
\left(1+\ln^2{\tau}\right)} & \scsz{\lfrac{1}{\left|\Upsilon\right|}
\tau\left(1+\ln^2{\tau}\right)} \\*[2mm]
\scsz{\dot{\phi}_3(t')} & \scsz{\left(1+\ln{\tau}\right)^2} & 
\scsz{-\left(1+\ln{\tau}\right)^2}\\*[2mm]
\scsz{\ddot{\phi}_3(t')} & \scsz{2\Upsilon\frac{1}{\tau}
\left(1+\ln{\tau}\right)} & \scsz{2\left|\Upsilon\right|\frac{1}{\tau}
\left(1+\ln{\tau}\right)}\\*[2mm]
\scsz{\phi_1(t')} & \scsz{\lfrac{1}{2\Upsilon}\tau\left(1-\ln^2{\tau}
+2i\ln{\tau}\right)} & \scsz{\lfrac{1}{2\left|\Upsilon\right|}\tau
\left(1-\ln^2{\tau}-2i\ln{\tau}\right)}\\*[2mm]
\scsz{\dot{\phi}_1(t')} & \scsz{\lfrac{1}{2}\left[1-\ln^2{\tau}-2\ln{\tau}
+2i\left(1+\ln{\tau}\right)\right]} & \scsz{-\lfrac{1}{2}\left[1-
\ln^2{\tau}-2\ln{\tau}-2i\left(1+\ln{\tau}\right)\right]}\\*[2mm]\hline
 & \multicolumn{2}{c} {$\Upsilon^2 < 4\omega^2$}\\*[1mm]\cline{2-3}
\scsz{\xi(t')} & \scsz{\sqrt{\lfrac{1}{\Upsilon\Delta}}\sqrt{\tau}
e^{i\lfrac{\Delta}{2}\ln{\tau}}} & \scsz{
\sqrt{\lfrac{1}{\left|\Upsilon\right|\Delta}}\sqrt{\tau}
e^{-i\lfrac{\Delta}{2}\ln{\tau}}}\\*[2mm]
\scsz{\dot{\xi}(t')} & \scsz{\sqrt{\lfrac{\Upsilon}{4\Delta}}
\left(1+i\Delta\right)\frac{1}{\sqrt{\tau}}e^{i\lfrac{\Delta}{2}
\ln{\tau}}} & -\scsz{\sqrt{\lfrac{\left|\Upsilon\right|}{4\Delta}}
\left(1-i\Delta\right)\frac{1}{\sqrt{\tau}}e^{-i\lfrac{\Delta}{2}
\ln{\tau}}}\\*[2mm]
\scsz{\phi_3(t')} & \scsz{\lfrac{2}{\Upsilon\Delta}\tau} & 
\scsz{\lfrac{2}{\left|\Upsilon\right|\Delta}\tau}\\*[2mm]
\scsz{\dot{\phi}_3(t')} & \scsz{\lfrac{2}{\Delta}} & 
\scsz{-\lfrac{2}{\Delta}}\\*[2mm]
\scsz{\ddot{\phi}_3(t')} & \scsz{0} & \scsz{0}\\*[2mm]
\scsz{\phi_1(t')} & \scsz{\lfrac{1}{\Upsilon\Delta}\tau e^{i\Delta
\ln{\tau}}} 
& \scsz{\lfrac{1}{\left|\Upsilon\right|\Delta}\tau e^{-i\Delta\ln{\tau}}}
\\*[2mm]
\scsz{\dot{\phi}_1(t')} & \scsz{\lfrac{1+i\Delta}{\Delta}
e^{i\Delta\ln{\tau}}} & \scsz{-\lfrac{1-i\Delta}{\Delta}
e^{-i\Delta\ln{\tau}}}\\*[2mm]\hline
\end{tabular}



\newpage

\begin{tabular}{c|c|c}
\multicolumn{3}{l} {Table B-2.  Time dependent functions 
for the $TM$ system. The}\\
\multicolumn{3}{l} {functions $\hat{\bar{\xi}}$, $\hat{\dot{\bar{\xi}}}$, 
$\phi_2$, and $\hat{\dot{\phi}}_2$ can be obtained by taking the 
complex conjugate of}\\
\multicolumn{3}{l} {$\hat{\xi}$, $\hat{\dot{\xi}}$, $\hat{\phi}_1$, 
and $\hat{\dot{\phi}}_1$, respectively. 
$\Delta^2=|1-4\omega^2/\Upsilon^2|$, $\chi=\Upsilon(t-t_o)$.}\\*[2mm]\hline
 & $\Upsilon > 0$ & $\Upsilon < 0$\\*[1mm]\hline
 & \multicolumn{2}{c} {$\Upsilon^2>4\omega^2$} \\*[1mm]\cline{2-3}
\scsz{\hat{\xi}(t)} & \scsz{\sqrt{\lfrac{1}{2\Upsilon\Delta}}e^{\chi/2}
\left(e^{-\lfrac{\Delta}{2}\chi}+ie^{\lfrac{\Delta}{2}\chi}
\right)} & \scsz{\sqrt{\lfrac{1}{2\left|\Upsilon\right|\Delta}}e^{\chi/2}
\left(e^{-\lfrac{\Delta}{2}\chi}-ie^{\lfrac{\Delta}{2}\chi}
\right)}  \\*[2mm]
\scsz{\hat{\dot{\xi}}(t)} & \scsz{\sqrt{\lfrac{\Upsilon}{8\Delta}}
e^{-\chi/2}\left[(1-\Delta)e^{-\lfrac{\Delta}{2}\chi}
+i(1+\Delta)e^{\lfrac{\Delta}{2}\chi}\right]} & 
\scsz{-\sqrt{\lfrac{\left|\Upsilon\right|}{8\Delta}}
e^{-\chi/2}\left[(1-\Delta)e^{-\lfrac{\Delta}{2}\chi}
-i(1+\Delta)e^{\lfrac{\Delta}{2}\chi}\right]}\\*[2mm]
\scsz{\hat{\phi}_3(t)} & \scsz{\lfrac{1}{\Upsilon\Delta}e^{\chi}\left(
e^{-\Delta\chi}+e^{\Delta\chi}\right)} & 
\scsz{\lfrac{1}{\left|\Upsilon\right|\Delta}e^{\chi}\left(
e^{-\Delta\chi}+e^{\Delta\chi}\right)}\\*[2mm]
\scsz{\hat{\dot{\phi}}_3(t)} & \scsz{\lfrac{1}{\Delta}\left[(1-\Delta)
e^{-\Delta\chi}+(1+\Delta)e^{\Delta\chi}\right]} &  
\scsz{-\lfrac{1}{\Delta}\left[(1-\Delta)
e^{-\Delta\chi}+(1+\Delta)e^{\Delta\chi}\right]} \\*[2mm]
\scsz{\hat{\ddot{\phi}}_3(t)} & \scsz{\Upsilon e^{-\chi}\left[-(1-\Delta)
e^{-\Delta\chi}+(1+\Delta)e^{\Delta\chi}\right]} & 
\scsz{\left|\Upsilon\right|e^{-\chi}\left[-(1-\Delta)
e^{-\Delta\chi}+(1+\Delta)e^{\Delta\chi}\right]} \\*[2mm]
\scsz{\hat{\phi}_1(t)} & \scsz{\lfrac{1}{2\Upsilon\Delta}e^{\chi}\left(
e^{-\Delta\chi}-e^{\Delta\chi}+2i\right)} & 
\scsz{\lfrac{1}{2\left|\Upsilon\right|\Delta}e^{\chi}\left(
e^{-\Delta\chi}-e^{\Delta\chi}-2i\right)}\\*[2mm]
\scsz{\hat{\dot{\phi}}_1(t)} & \scsz{\lfrac{1}{2\Delta}\left[(1-\Delta)
e^{-\Delta\chi}-(1+\Delta)e^{\Delta\chi}+2i\right]} & 
\scsz{-\lfrac{1}{2\Delta}\left[(1-\Delta)
e^{-\Delta\chi}-(1+\Delta)e^{\Delta\chi}-2i\right]} \\*[2mm]\hline
 & \multicolumn{2}{c} {$\Upsilon^2 = 4\omega^2$} \\*[1mm]\cline{2-3}
\scsz{\hat{\xi}(t)} & \scsz{\sqrt{\lfrac{1}{2\Upsilon}}e^{\chi/2}
\left(1+i\chi\right)} & \scsz{\sqrt{\lfrac{1}{2\left|\Upsilon\right|}}
e^{\chi/2}\left(1-i\chi\right)}\\*[2mm]
\scsz{\hat{\dot{\xi}}(t)} & \scsz{\sqrt{\lfrac{\Upsilon}{2}}
e^{-\chi/2}\left[\lfrac{1}{2}+i\left(1+\lfrac{1}{2}
\chi\right)\right]} & \scsz{-\sqrt{\lfrac{\left|\Upsilon\right|}{2}}
e^{-\chi/2}\left[\lfrac{1}{2}-i\left(1+\lfrac{1}{2}
\chi\right)\right]}\\*[2mm]
\scsz{\hat{\phi}_3(t)} & \scsz{\lfrac{1}{\Upsilon}e^{\chi}
\left(1+\chi^2\right)} & \scsz{\lfrac{1}{\left|\Upsilon\right|}
e^{\chi}\left(1+\chi^2\right)} \\*[2mm]
\scsz{\hat{\dot{\phi}}_3(t)} & \scsz{\left(1+\chi\right)^2} & 
\scsz{-\left(1+\chi\right)^2}\\*[2mm]
\scsz{\hat{\ddot{\phi}}_3(t)} & \scsz{2\Upsilon e^{-\chi}
\left(1+\chi\right)} & \scsz{2\left|\Upsilon\right|e^{-\chi}
\left(1+\chi\right)}\\*[2mm]
\scsz{\hat{\phi}_1(t)} & \scsz{\lfrac{1}{2\Upsilon}e^{\chi}\left(1-\chi^2
+2i\chi\right)} & \scsz{\lfrac{1}{2\left|\Upsilon\right|}e^{\chi}
\left(1-\chi^2-2i\chi\right)}\\*[2mm]
\scsz{\hat{\dot{\phi}}_1(t)} & \scsz{\lfrac{1}{2}\left[1-\chi^2-2\chi
+2i\left(1+\chi\right)\right]} & \scsz{-\lfrac{1}{2}\left[1-
\chi^2-2\chi-2i\left(1+\chi\right)\right]}\\*[2mm]\hline
 & \multicolumn{2}{c} {$\Upsilon^2 < 4\omega^2$}\\*[1mm]\cline{2-3}
\scsz{\hat{\xi}(t)} & \scsz{\sqrt{\lfrac{1}{\Upsilon\Delta}}e^{\chi/2}
e^{i\lfrac{\Delta}{2}\chi}} & \scsz{
\sqrt{\lfrac{1}{\left|\Upsilon\right|\Delta}}e^{\chi/2}
e^{-i\lfrac{\Delta}{2}\chi}}\\*[2mm]
\scsz{\hat{\dot{\xi}}(t)} & \scsz{\sqrt{\lfrac{\Upsilon}{4\Delta}}
\left(1+i\Delta\right)e^{-\chi/2}e^{i\lfrac{\Delta}{2}
\chi}} & -\scsz{\sqrt{\lfrac{\left|\Upsilon\right|}{4\Delta}}
\left(1-i\Delta\right)e^{-\chi/2}e^{-i\lfrac{\Delta}{2}\chi}}\\*[2mm]
\scsz{\hat{\phi}_3(t)} & \scsz{\lfrac{2}{\Upsilon\Delta}e^{\chi}} & 
\scsz{\lfrac{2}{\left|\Upsilon\right|\Delta}e^{\chi}}\\*[2mm]
\scsz{\hat{\dot{\phi}}_3(t)} & \scsz{\lfrac{2}{\Delta}} & 
\scsz{-\lfrac{2}{\Delta}}\\*[2mm]
\scsz{\hat{\ddot{\phi}}_3(t)} & \scsz{0} & \scsz{0}\\*[2mm]
\scsz{\hat{\phi}_1(t)} & \scsz{\lfrac{1}{\Upsilon\Delta}e^{\chi} 
e^{i\Delta\chi}} 
& \scsz{\lfrac{1}{\left|\Upsilon\right|\Delta}e^{\chi} e^{-i\Delta\chi}}
\\*[2mm]
\scsz{\hat{\dot{\phi}}_1(t)} & \scsz{\lfrac{1+i\Delta}{\Delta}
e^{i\Delta\chi}} & \scsz{-\lfrac{1-i\Delta}{\Delta}
e^{-i\Delta\chi}}\\*[2mm]\hline
\end{tabular}


\newpage

\begin{tabular}{c|c|c}
\multicolumn{3}{l} {Table B-3.  Time dependent functions 
for the $TQ$ system. The}\\
\multicolumn{3}{l} {functions $\bar{\Xi}_P$ and $\bar{\Xi}_X$ can be 
obtained by taking the complex conjugate of}\\
\multicolumn{3}{l} {$\Xi_P$ and  $\Xi_X$, respectively. 
$\Delta^2=|1-4\omega^2/\Upsilon^2|$, $\chi=\Upsilon(t-t_o)$.}
\\*[2mm]\hline
 & $\Upsilon > 0$ & $\Upsilon < 0$\\*[1mm]\hline
 & \multicolumn{2}{c} {$\Upsilon^2>4\omega^2$} \\*[1mm]\cline{2-3}
\scsz{\Xi_P(t)} & \scsz{\sqrt{\lfrac{1}{2\Upsilon\Delta}}
\left(e^{-\lfrac{\Delta}{2}\chi}+ie^{\lfrac{\Delta}{2}\chi}
\right)} & \scsz{\sqrt{\lfrac{1}{2\left|\Upsilon\right|\Delta}}
\left(e^{-\lfrac{\Delta}{2}\chi}-ie^{\lfrac{\Delta}{2}\chi}
\right)}  \\*[2mm]
\scsz{\Xi_X(t)} & \scsz{\sqrt{\lfrac{\Upsilon}{8\Delta}}
\left[(1-\Delta)e^{-\lfrac{\Delta}{2}\chi}
+i(1+\Delta)e^{\lfrac{\Delta}{2}\chi}\right]} & 
\scsz{-\sqrt{\lfrac{\left|\Upsilon\right|}{8\Delta}}
\left[(1-\Delta)e^{-\lfrac{\Delta}{2}\chi}
-i(1+\Delta)e^{\lfrac{\Delta}{2}\chi}\right]}\\*[2mm]
\scsz{C_{3,T}(t)} & \scsz{\lfrac{1}{\Upsilon\Delta}\left(
e^{-\Delta\chi}+e^{\Delta\chi}\right)} & 
\scsz{\lfrac{1}{\left|\Upsilon\right|\Delta}\left(
e^{-\Delta\chi}+e^{\Delta\chi}\right)}\\*[2mm]
\scsz{C_{3,D}(t)} & \scsz{\lfrac{1}{2}\left[-
e^{-\Delta\chi}+e^{\Delta\chi}\right]} &  
\scsz{\lfrac{1}{2}\left[
e^{-\Delta\chi}-e^{\Delta\chi}\right]} \\*[2mm]
\scsz{C_{3,X^2}(t)} & \scsz{-\lfrac{\Upsilon}{4}\left[-(1-\Delta)
e^{-\Delta\chi}+(1+\Delta)e^{\Delta\chi}\right]} & 
\scsz{\lfrac{-\left|\Upsilon\right|}{4}\left[-(1-\Delta)
e^{-\Delta\chi}+(1+\Delta)e^{\Delta\chi}\right]} \\*[2mm]\hline
 & \multicolumn{2}{c} {$\Upsilon^2 = 4\omega^2$} \\*[1mm]\cline{2-3}
\scsz{\Xi_P(t)} & \scsz{\sqrt{\lfrac{1}{2\Upsilon}}
\left(1+i\chi\right)} & \scsz{\sqrt{\lfrac{1}{2\left|\Upsilon\right|}}
\left(1-i\chi\right)}\\*[2mm]
\scsz{\Xi_X(t)} & \scsz{\sqrt{\lfrac{\Upsilon}{2}}
\left[\lfrac{1}{2}+i\left(1+\lfrac{1}{2}
\chi\right)\right]} & \scsz{-\sqrt{\lfrac{\left|\Upsilon\right|}{2}}
\left[\lfrac{1}{2}-i\left(1+\lfrac{1}{2}\chi\right)\right]}\\*[2mm]
\scsz{C_{3,T}(t)} & \scsz{\lfrac{1}{\Upsilon}
\left(1+\chi^2\right)} & \scsz{\lfrac{1}{\left|\Upsilon\right|}
\left(1+\chi^2\right)} \\*[2mm]
\scsz{C_{3,D}(t)} & \scsz{\chi} & \scsz{-\chi}\\*[2mm]
\scsz{C_{3,X^2}(t)} & \scsz{-\lfrac{\Upsilon}{2} 
\left(1+\chi\right)} & \scsz{-\lfrac{\left|\Upsilon\right|}{2}
\left(1+\chi\right)}\\*[2mm]\hline
 & \multicolumn{2}{c} {$\Upsilon^2 < 4\omega^2$}\\*[1mm]\cline{2-3}
\scsz{\Xi_P(t)} & \scsz{\sqrt{\lfrac{1}{\Upsilon\Delta}}
e^{i\lfrac{\Delta}{2}\chi}} & \scsz{
\sqrt{\lfrac{1}{\left|\Upsilon\right|\Delta}}
e^{-i\lfrac{\Delta}{2}\chi}}\\*[2mm]
\scsz{\Xi_X(t)} & \scsz{\sqrt{\lfrac{\Upsilon}{4\Delta}}
\left(1+i\Delta\right)e^{i\lfrac{\Delta}{2}
\chi}} & -\scsz{\sqrt{\lfrac{\left|\Upsilon\right|}{4\Delta}}
\left(1-i\Delta\right)e^{-i\lfrac{\Delta}{2}\chi}}\\*[2mm]
\scsz{C_{3,T}(t)} & \scsz{\lfrac{2}{\Upsilon\Delta}} & 
\scsz{\lfrac{2}{\left|\Upsilon\right|\Delta}}\\*[2mm]
\scsz{C_{3,D}(t)} & \scsz{0} & \scsz{0}\\*[2mm]
\scsz{C_{3,X^2}(t)} & \scsz{0} & \scsz{0}\\*[2mm]\hline
\end{tabular}



\newpage

\baselineskip=.27in

\end{document}